\documentclass[aps,pre,superscriptaddress,floatfix,tightenlines,showpacs,twocolumn,10pt,notitlepage]{revtex4-2}
\usepackage{amsmath}
\usepackage{amssymb}
\usepackage{amsthm}
\usepackage{bm}
\usepackage{bbm}
\usepackage[shortlabels]{enumitem}
\usepackage{tabularx}
\usepackage{graphicx}
\usepackage{tikz-cd}
\usepackage{xcolor}
\usepackage{hyperref}
\usepackage[normalem]{ulem}
\usepackage[textsize=scriptsize,backgroundcolor=white]{todonotes}
\usepackage{url}
\usepackage[dvipsnames,svgnames,table]{xcolor}
\usepackage{framed}
\usepackage{color}
\usepackage{comment}
\usepackage{mathtools}
\usepackage{booktabs, multirow}
\usepackage{subcaption}
\usepackage[ruled,vlined]{algorithm2e}

\hypersetup{
	colorlinks=true,
	linkcolor=blue,
	citecolor=blue,			
	urlcolor=blue,			
	filecolor=blue,
	anchorcolor = green,
}

\begin{document}
\title{Variational quantum algorithm for anion exchange across electrolyzer membrane}

\author{Timur Gubaev}
\affiliation{Institute of Thermodynamics and Fluid Mechanics, Technische Universit\"at Ilmenau, P.O.\ Box 100565, D-98684 Ilmenau, Germany}

\author{Philipp Pfeffer}
\affiliation{Institute of Thermodynamics and Fluid Mechanics, Technische Universit\"at Ilmenau, P.O.\ Box 100565, D-98684 Ilmenau, Germany}

\author{Christian Dreßler}
\affiliation{Institute of Physics, Technische Universit\"at Ilmenau, P.O.\ Box 100565, D-98684 Ilmenau, Germany}

\author{Jörg Schumacher}
\affiliation{Institute of Thermodynamics and Fluid Mechanics, Technische Universit\"at Ilmenau, P.O.\ Box 100565, D-98684 Ilmenau, Germany}

\date{\today}

\begin{abstract}
We present a variational quantum algorithm that solves the one-dimensional diffusion problem with a space-dependent diffusion constant $D(x)$. This problem is relevant for the exchange of hydroxide ions across a two-layer membrane in an alkaline electrolyzer, where the concentration of OH$^-$ ion determines the chemical stability for longer time periods. We use $16$ to $64$ grid points across the membrane, resulting from  $n=4$ to 6 data qubits for the ideal statevector and shot-based quantum simulations implemented using Qiskit. For these qubit numbers, the depth of the parametric quantum circuit has been chosen to ensure sufficient expressibility. The state preparation requires particular attention since the diffusivity $D$ is piecewise constant in the different layers with discontinuities at the interface. Furthermore, we compare different classical optimization schemes with respect to their convergence in the VQA method. We demonstrate the applicability of the quantum algorithm to a problem with non-trivial boundary conditions and jump conditions of the diffusion constant and outline possible extensions of the proof-of-concept application case of quantum computing. Our simulations show that pronounced hydroxide ion concentration gradients, and thus chemical instabilities, can occur only when the ratio of diffusivity in both layers of the membrane exceeds approximately 50.
\end{abstract}

\maketitle

\section{Introduction}
Electrolyzers and fuel cells are a central component for enabling the transformation to a green economy, which use electricity, e.g. from renewable energy, to split water into oxygen and hydrogen --- a clean fuel and carrier of energy \cite{Yan2020,Elshafie2023}. Membranes and diaphragms are key components of alkaline electrolyzers, where they separate the anode and cathode compartments while allowing ionic transport. Conventional alkaline water electrolysis (AWE) employs porous diaphragms such as Zirfon (ZrO$_2$--polysulfone), valued for their mechanical stability, chemical durability, and low area resistance ($\sim$0.1~$\Omega$\,cm$^2$ at 80~$^\circ$C in 30~wt\% KOH)~\cite{Lee2020}. However, their relatively high gas crossover, especially under pressure gradients between cathode and anode, limits efficiency and safety in pressurized operation~\cite{Schalenbach2016,Brauns2021}. In contrast, anion exchange membranes (AEMs) offer low H$_2$/O$_2$ crossover and enable differential pressure operation~\cite{Delp2025,Du2022}, but suffer from limited mechanical robustness and higher cost. Recent reviews highlight the potential of hybrid or graded separators that combine the dense, selective character of AEMs with the mechanical stability of porous supports (multi-layered AEM)~\cite{Henkensmeier2024,Sugawara2023}. Such configurations will be studied in the following. 

A major drawback of anion-exchange membranes (AEMs) remains their relatively low ionic conductivity and their limited long-term stability in alkaline environments \cite{yassnin2025,mustain2025}. The issue of low conductivity can be mitigated in hybrid membrane architectures, in which a very thin AEM layer provides low area resistance, while a mechanically robust porous support layer ensures structural integrity. The present study focuses on the second issue, namely alkaline stability. It is well established that AEMs exhibit markedly higher chemical stability under well-controlled and fully hydrated laboratory conditions, whereas their durability decreases substantially under realistic electrolyzer operation with enhanced pressure \cite{diesendruck2018,koch2022}. This reduction in stability is closely linked to local gradients in the hydroxide (OH) concentration, which become particularly pronounced when water management is insufficient and membrane hydration is low.

In this study, we examine whether discontinuities in the effective diffusion coefficient, which arise from combining two different materials  with different transport properties in a hybrid membrane, can lead to critically high hydroxide concentration gradients under operational conditions. These gradients are of central importance because they are directly related to chemical degradation processes within the membrane, ionomer stability, and ultimately the lifetime and performance of the electrolyzer.

 To this end, the transport of the negatively charged anions (OH$^{-}$) across the membrane is macroscopically modeled as a migration process with a space-dependent diffusion constant, see e.g. \cite{Marino2014}. The numerical modeling of these transport processes is established by a broad spectrum of methods ranging from microscopic molecular dynamics (MD) models \cite{zelovich2019,chen2016, otmi2024} to primarily one- and two-dimensional diffusion processes with or without additional drift terms, which are for example caused by electrical fields \cite{Bard2001}. While the microscopic models describe the short-term processes, the macrospcopic diffusion provides insights on the longer-term behavior, and thus stability. Particularly, one-dimensional transport equations can be solved these days by quantum computing method using a small number of qubits.

The solution of classical physics problems using quantum computing \cite{Nielsen2010,Preskill2018} is an emerging research field, which explores the potential capabilities of quantum devices and systems to study dynamics that is often formulated in the form of partial differential equations (PDE), such as in fluid mechanics \cite{Bharadwaj2020,Succi2023,Tennie2025}. One class of algorithms to investigate transport processes are variational quantum algorithms (VQA). When applied to solving differential equations, VQAs typically reformulate the problem as an optimization task; they are hybrid quantum-classical algorithms \cite{Cerezo2021}. A popular field of application are linear one-dimensional advection-diffusion equations for simple transport problems \cite{Demirdjian2022,Leong2022,Leong2023,Jaksch2023,Ingelmann2024,bengoechea2025towards}. The framework was extended from frequently used periodic \cite{Ingelmann2024} to Dirichlet and Neumann boundary conditions in ref. \cite{over2025boundary}. Furthermore, linear heat equations in one and two dimensions were considered by VQAs in refs. \cite{Guseynov2023,Liu2023}. VQA applications to nonlinear partial differential equations include steady \cite{Lubasch2020} and time-dependent \cite{Koecher2025} one-dimensional nonlinear Schrödinger equations, as well as the one-dimensional Burgers equation \cite{Lubasch2020,Pool2022,Pool2024}. In refs. \cite{Pool2022,Pool2024} an alternative Feynman-Kitaev algorithm is used which orders spatial and temporal qubits in one register and thus avoids time stepping. 

In the present work, we investigate the diffusive anion transport problem across a multi-layered AEM by a variational quantum algorithm in ideal statevector or shot-based quantum simulation framework. The studied classical field is the anion concentration $c(x,t)$ for $t>0$ and $x\in [\tilde{x}_0,\tilde{x}_m]$. In contrast to a recent quantum spectral method for constant diffusion~\cite{Pfeffer2025}, our application is characterized by a linear PDE with piecewise constant diffusion constants $D(x)$ in the different layers of the AEM,
\begin{equation}
\frac{\partial c}{\partial t} = {\cal L}[c(x,t), D(x)]=\frac{\partial}{\partial x}\left(D(x) \frac{\partial c}{\partial x}\right)\,,
\label{eq:diffusion_equation}
\end{equation}
where ${\cal L}$ is a linear operator. We apply different Dirichlet boundary conditions at the two ends of the membrane cross-section, which have to be included in the variational formulation. We derive an analytical solution for the temporal relaxation to a steady ion concentration, which allows us to validate the quantum algorithm for different grid resolutions. The latter are connected to the number of qubits: a discretization of the interval into $N$ equidistant mesh cells requires $n=\log_2(N)$ qubits. The state preparation is adapted here to capture for the discontinuities of $D$ in the multi-layered AEM. In addition, we compare different optimization algorithms for the problem at hand, among them a recently suggested surrogate model-based optimization scheme \cite{Shaffer2023} for the search of the minimum of nonconvex cost functions. This optimization step of the VQA typically complicates complexity estimates of the algorithm as the optimization procedure can run into barren plateaus \cite{Holmes2022}. Furthermore, we investigate the expressivity of the quantum ansatz for different depths \cite{sim2019expressibility}. Our proof-of-concept study demonstrates the capability of VQA algorithms to solve the concrete application case at hand. We discuss possible straight extensions of the anion transport problem.

The hybrid quantum-classical nature of VQAs implies that the cost function $C({\bm \lambda})$ is minimized classically by an optimization method \cite{Cerezo2021}. This cost function is evaluated by a parametric quantum circuit, which is composed of $n$ qubits and single- and two-qubit gates. The cost minimum search is performed classically in a high-dimensional parameter space of dimension $M\sim {\cal O}(N)$ that contains a parameter vector ${\bm \lambda}$. This parameter vector ${\bm \lambda}$, which consists of the angles of the single-qubit unitary rotation gates ${\bm \lambda} = (\lambda_1, \lambda_2, \dots, \lambda_M)\in \mathbb{R}^M$, is the input to the algorithm. Several approaches exist for the implementation, such as minimizing the Euclidean distance between the VQA and a finite difference solution \cite{Ingelmann2024}, or adopting a weak variational formulation \cite{bengoechea2025towards, over2025boundary}, the latter of which will be used here (and detailed below). 

The outline is as follows. In Sec. II, we explain the classical framework including the analytical solution of the initial-boundary value problem, the relaxation rate to the steady state profile, and the finite difference method which will be used for the comparison with the quantum algorithm. Section III discusses the specifics of the variational quantum algorithm including the quantum state preparation. In Sec. IV, the results and analyses are presented. This includes expressibility study, a comparison of classical optimization algorithms, and a detailed discussion of the time evolution towards the steady solution in comparison to the classical finite difference method. We conclude with a summary and an outlook. 

\section{Classical Framework}

\subsection{Problem description and analytical solution}
\begin{figure*}
    \centering
    \includegraphics[width=0.45\textwidth]{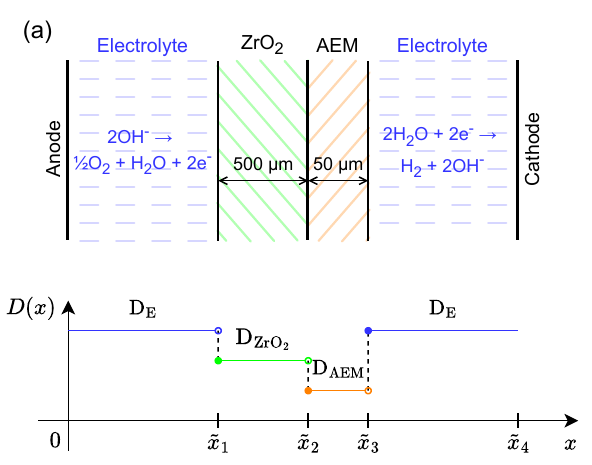}
    \includegraphics[width=0.45\textwidth]{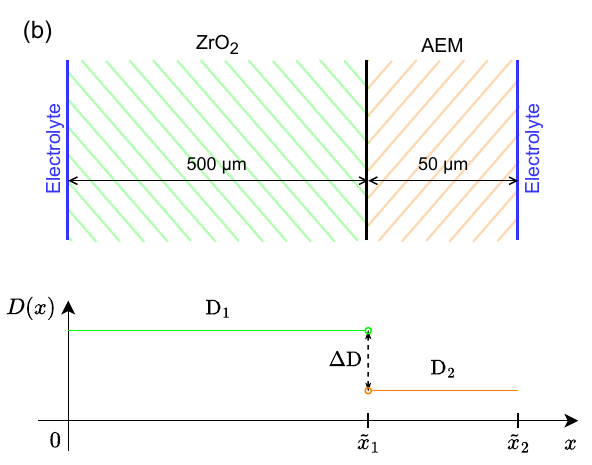}
    \caption{Sketch of the anion-exchange membrane (AEM). (a) The 4-layer AEM electrolysis system. (b) The simplified configuration, which will be used for the actual quantum computation approximates the two electrolyte reservoirs by Dirichlet conditions for the anion concentration at the corresponding outer membrane interfaces. The layer $\text{Zr}\text{O}_2$ refers to Zirfon, a composite material consisting of $\text{Zr}\text{O}_2$ embedded in a polysulfone matrix. It serves as a stabilizer.}
    \label{fig:aem_problem}
\end{figure*}

We employ a variational quantum algorithm (VQA) to numerically simulate hydroxide-ion transport in alkaline water electrolysis. This diffusive transport process with spatially varying diffusivity $D$ is depicted in Fig.~\ref{fig:aem_problem}(a). The system produces hydrogen gas at the cathode by reducing water molecules, generating hydroxide ions as a byproduct. These hydroxide ions migrate through the anion exchange membrane (AEM) to the anode, where they are oxidized to produce oxygen gas and release electrons. The process efficiently converts electrical energy to chemical energy stored in hydrogen, while safely separating hydrogen and oxygen production.  The number of layers in the original AEM electrolysis problem can be reduced to two layers, see Fig.~\ref{fig:aem_problem}(b), by assuming that the electrolytes are well-mixed, such that the charge distribution within them remains uniform. 

The anion-exchange membrane electrolysis problem is described by the one-dimensional diffusion equation \eqref{eq:diffusion_equation} with piecewise-constant diffusivity in $\Omega_T = (\tilde{x}_0, \tilde{x}_m) \times (0, T]$. The following initial condition,
\begin{equation}
    \label{eq:initial_conditions}
    c(x, 0) = c_0(x)\,,
\end{equation}
and Dirichlet boundary conditions do hold,
\begin{equation}
    \label{eq:boundary_conditions}
    c(\tilde{x}_0, t) = c_{0, 1}\quad \mbox{and}\quad
    c(\tilde{x}_m, t) = c_{0, m}\,,
\end{equation}
where $\tilde{x}_0 = 0\,\text{m}$ and $\tilde{x}_m = 10^{-4}\,\textrm{m}$ are interval boundaries; $\ell=\tilde{x}_m-\tilde{x}_0 = 10^{-4}\,\textrm{m}$ is a domain width of sub-millimeter size which is typical of these problems, $m$ is the number of layers, $T$ is the time horizon, $c_0(x)$ is the initial condition at location $x$, $c_{0, 1}$ is the concentration at the left boundary and $c_{0, m}$ is the concentration at the right boundary. While $\tilde x_0$ and $\tilde x_m$ are the boundary points, locations $\tilde x_1, \tilde x_2, \dots, \tilde x_{m - 1}$ represent the interfaces between the layers. For example, the number of layers is $m = 4$ for the problem in Fig.~\ref{fig:aem_problem}(a) with interfaces $\tilde x_1$, $\tilde x_2$ and $\tilde x_3$ and $m = 2$ for the problem in Fig.~\ref{fig:aem_problem}(b) with a single interface $\tilde x_1$, respectively. Interfaces are subject to interface conditions discussed in the next section.

In eqns. \eqref{eq:diffusion_equation} and \eqref{eq:initial_conditions}, the diffusion coefficient $D(x)$ and the initial conditions $c_0(x)$ are piecewise-constant functions defined as
\begin{align}
    \label{eq:diffusion_coefficient}
    D(x) &= D_j \,,\\
   \label{eq:initial_conditions_function}
    c_0(x) &= c_{0, j} \,, 
\end{align}
for $\tilde{x}_{j - 1} \leq x < \tilde{x}_j$ and $j = 1, \dots, m$. Note that at the interface positions $\tilde{x}_j$ the diffusion coefficient and the initial concentration is taken from the region to the right of the interface. In eq. \eqref{eq:diffusion_coefficient}, the diffusivity of the Zirfon layer is $D_1\sim10^{-8}\,\text{m}^2\,/\text{s}$, whereas the diffusivity of the AEM, $D_2$, will be varied between $10^{-8}\,\text{m}^2/\text{s}$ and $10^{-10}\,\text{m}^2/\text{s}$ in subsequent numerical experiments.

To non-dimensionalize the problem, we define the characteristic time scale $\tau_D$ in terms of the domain length and the Zirfon layer diffusivity as $\tau_D = \ell^2/D_1$. The variables $x$, $\tilde x_j$, $t$ and $D_j$ are then non-dimensionalized as
\begin{align}
    \bar x = \frac{x}{\ell} \,,\quad
    \bar{\tilde x}_j = \frac{\tilde x_j}{\ell} \,,\quad
    \bar t = \frac{t}{\tau_D} \,,\quad \text{and} \quad
    \bar D_j = \frac{D_j\tau_D}{\ell^2} \,. 
\end{align}
To keep notation simple, we omit from now on the overbars and re-use $x$, $\tilde x_j$, $t$ and $D_j$ to denote the corresponding non-dimensional variables $\bar x$, $\bar{\tilde x}_j$, $\bar t$ and $\bar D_j$. In the following sections, the steady state and analytical solutions will be derived for an arbitrary number of layers $m \geq 2$.

\subsubsection{Steady state analytical solution}
To obtain a steady state solution in the presence of discontinuities, the right-hand side of \eqref{eq:diffusion_equation} is equated to zero in $m$ individual regions $(\tilde{x}_{j - 1}, \tilde{x}_j)$ with each region corresponding to the $j$-th layer, and, additionally, interface conditions are imposed at the interfaces $\tilde x_1, \tilde x_2, \dots, \tilde x_{m - 1}$ between the layers where discontinuities arise. When a region $(\tilde{x}_{j - 1}, \tilde{x}_j)$ is considered, the equation \eqref{eq:diffusion_equation} simplifies as,
\begin{align*}
    \frac{\partial c}{\partial t} = \frac{\partial}{\partial x}\left(D(x)\frac{\partial c}{\partial x}\right) \quad \Rightarrow \quad \frac{\partial c}{\partial t} = D_j\frac{\partial^2 c}{\partial x^2}\,,
\end{align*}
where $D_j$ is the diffusivity of the $j$-th layer. Then, the right-hand side of the simplified equation is equated to zero while treating $c(x, t)$ as the steady state solution $c_s(x)$, and an integral is taken with respect to $x$,
\begin{align*}
    D_j\frac{d^2 c_s}{dx^2} = 0 \quad \Rightarrow \quad  D_j\frac{dc_s}{dx} = J_j\,,
\end{align*}
where $J_j$ is the flux in the $j$-th layer. Taking another integral yields the steady state solution for $\tilde{x}_{j - 1} \leq x < \tilde{x}_j$ and $j = 1, \dots, m$,
\begin{align*}
    c_s(x) = \frac{J_j}{D_j}x + b_j\,,
\end{align*}
where $b_j$ is an integration constant. Next, to further address the discontinuities and determine the unknowns $J_j$ and $b_j$, the interface conditions are imposed at the interface positions $\tilde{x}_j$ as $j = 1, 2, \dots, m - 1$ along with boundary conditions, namely
\begin{align}
    \label{eq:steady_state_solution_continuity}
    \lim_{x \rightarrow \tilde{x}_j^-} c_s(x) &= \lim_{x \rightarrow \tilde{x}_j^+} c_s(x)
    \,,\\
    \label{eq:steady_state_flux_continuity}
    \lim_{x \rightarrow \tilde{x}_j^-} D(x) \frac{dc_s}{dx} &= \lim_{x \rightarrow \tilde{x}_j^+} D(x)\frac{dc_s}{dx}
    \,,\\    
    \label{eq:steady_state_boundary_conditions}
    c_s(\tilde{x}_0) = c_{0, 1} \quad &\text{and} \quad c_s(\tilde{x}_m) = c_{0, m}\,.
\end{align}
Equations \eqref{eq:steady_state_solution_continuity} and \eqref{eq:steady_state_flux_continuity} correspond to solution and flux continuity, respectively, at interface positions. Eq. \eqref{eq:steady_state_flux_continuity} is satisfied when $J_j = J_{j + 1}$, meaning that fluxes in all $m$ layers are equal and can be equated to a global flux $J$, thus leaving only $m$ unknown constants $b_j$ and one unknown flux $J$,
\begin{align*}
    J_1 = J_2 = \dots = J_m = J\,.
\end{align*}
Substituting for $c_s$ in eqns. \eqref{eq:steady_state_solution_continuity} and \eqref{eq:steady_state_boundary_conditions}, and using the global flux $J$ results in the following linear system:
\begin{align}
    \label{eq:steady_state_solution_continuity_explicit}
    \frac{J}{D_j}\tilde{x}_j + b_j &= \frac{J}{D_{j + 1}}\tilde{x}_j + b_{j + 1}\,,\\
    \label{eq:steady_state_left_boundary_explicit}
    \frac{J}{D_1}\tilde{x}_0 + b_1 &= c_{0, 1}\,,\\
    \label{eq:steady_state_right_boundary_explicit}
    \frac{J}{D_m}\tilde{x}_m + b_m &= c_{0, m}\,.
\end{align}
Equations \eqref{eq:steady_state_solution_continuity_explicit} and \eqref{eq:steady_state_left_boundary_explicit} can be rewritten as a recurrence relation of $b_j$ in terms of $J$ as:
\begin{align}
    \label{eq:b_j_base}
    b_1 &= c_{0, 1} - \frac{J}{D_1}\tilde{x}_0\,,\\
   \label{eq:b_j_step}
    b_j &= b_{j - 1} + J\left(D_{j - 1}^{-1} - D_j^{-1}\right)\tilde{x}_{j - 1}\,, 
\end{align}
for $j = 2, 3, \dots, m$. Unrolling eqns. \eqref{eq:b_j_base} and \eqref{eq:b_j_step} produces a non-recursive closed-form formula for arbitrary $b_j$ in terms of the flux $J$:
\begin{equation}
    \label{eq:b_j}
    b_j = J\left[\sum_{k = 2}^j\left[\left(D_{k - 1}^{-1} - D_k^{-1}\right)\tilde{x}_{k - 1}\right] - D_1^{-1}\tilde{x}_0\right] + c_{0, 1}
\end{equation}
Using \eqref{eq:b_j} to evaluate $b_m$ and combining it with \eqref{eq:steady_state_right_boundary_explicit} produces an expression for $J$ which is given by
\begin{equation}
    \label{eq:J}
        J = \dfrac{c_{0, m} - c_{0, 1}}{D_m^{-1}\tilde{x}_m
         + \sum\limits_{k = 2}^m\left[\left(D_{k - 1}^{-1} - D_k^{-1}\right)\tilde{x}_{k - 1}\right] - D_1^{-1}\tilde{x}_0}\,.
\end{equation}
To simplify the expression for $c_s$ and further derivations, the prefactor of $x$ in $c_s(x)$ is defined as:
\begin{equation}
    \label{eq:a_j}
    a_j = \frac{J}{D_j}\,.
\end{equation}
The steady state solution is then expressed as
\begin{equation}
    \label{eq:steady_state}
    c_s(x) = a_j x + b_j, \quad \tilde{x}_{j - 1} \leq x < \tilde{x}_j, \;\; j = 1, \dots, m\,.
\end{equation}
To calculate the steady state solution the flux $J$ is obtained first using \eqref{eq:J}. Then $a_j$ are computed using \eqref{eq:a_j}, and $b_j$ are computed using recurrence relation \eqref{eq:b_j_base} and \eqref{eq:b_j_step}.

\subsubsection{Full time-dependent analytical solution}
Similarly to the heat equation with uniform thermal diffusivity, eq. \eqref{eq:diffusion_equation} is solved by the method of separation of variables. We use the ansatz
\begin{equation}
    \label{eq:separation_of_variables}
    c(x, t) = K(x)T(t)\,.
\end{equation}
In each region $(\tilde{x}_{j - 1}, \tilde{x}_j)$, \eqref{eq:diffusion_equation} simplifies as:
\begin{equation}
    \label{eq:diffusion_equation_j}
    \frac{\partial c}{\partial t} = D_j\frac{\partial^2 c}{\partial x^2} \quad \text{in} \quad \Omega_T = (\tilde{x}_{j - 1}, \tilde{x}_j) \times (0, T]\,.
\end{equation}
Substituting \eqref{eq:separation_of_variables} into \eqref{eq:diffusion_equation_j} and rearranging terms gives
\begin{equation}
    \label{eq:diffusion_equation_separated_j}
    \frac{1}{T(t)}\frac{dT}{dt} = \frac{D_j}{K_j(x)}\frac{d^2K_j}{dx^2}\,.
\end{equation}
Since left-hand and right-hand sides of \eqref{eq:diffusion_equation_separated_j} have to be equal for all $x$ and $t$, they must be equal to some separation constant which can be written as $-\lambda_k^2$. This gives rise to two ordinary differential equations for each $\lambda_k^2$:
\begin{align}
    \label{eq:ode_T_k}
    \frac{dT_k}{dt} &+ \lambda_k^2 T_k(t) = 0\,,\\
    \label{eq:ode_K_jk}
    \frac{d^2K_{jk}}{dx^2} &+ \frac{\lambda_k^2}{D_j}K_{jk}(x) = 0\,.
\end{align}
The solution to \eqref{eq:ode_T_k} reads $T_k(t) = C_k\exp\left(-\lambda_k^2 t\right)$. However, since $c(x, t) = K(x)T(t)$ the constant $C_k$ can be absorbed by the constants appearing in the solution of \eqref{eq:ode_K_jk}. Therefore, \eqref{eq:ode_T_k} and \eqref{eq:ode_K_jk} admit the following solutions:
\begin{align}
    \label{eq:T_k}
    T_k(t) &= \exp(-\lambda_k^2 t)\,,\\
    \label{eq:K_jk_intermediate}
    \begin{split}
        K_{jk}(x) &= \tilde{A}_{jk}\exp\left(-i\frac{\lambda_k}{\sqrt{D_j}}x\right) \\
        &+ \tilde{B}_{jk}\exp\left(i\frac{\lambda_k}{\sqrt{D_j}}x\right)\,.
    \end{split}
\end{align}
To simplify the further derivation, the following substitutions for constants $A_{jk}$ and $B_{jk}$ are introduced such that:
\begin{align}
    \label{eq:tilde_A_jk}
    \begin{split}
        \tilde{A}_{jk} &= \frac{1}{2}\Bigg[A_{jk}\exp\left(i\frac{\lambda_k}{\sqrt{D_j}}\tilde{x}_{j - 1}\right) \\
        &+ iB_{jk}\exp\left(i\frac{\lambda_k}{\sqrt{D_j}}\tilde{x}_{j - 1}\right)\Bigg]\,,
    \end{split}\\
   \label{eq:tilde_B_jk}
    \begin{split}
        \tilde{B}_{jk} &= \frac{1}{2}\Bigg[A_{jk}\exp\left(-i\frac{\lambda_k}{\sqrt{D_j}}\tilde{x}_{j - 1}\right) \\
        &- iB_{jk}\exp\left(-i\frac{\lambda_k}{\sqrt{D_j}}\tilde{x}_{j - 1}\right)\Bigg]\,.
    \end{split}
\end{align}
Substituting eqns. \eqref{eq:tilde_A_jk} and \eqref{eq:tilde_B_jk} into eq. \eqref{eq:K_jk_intermediate} yields the following expression in terms of the newly introduced coefficients $A_{jk}$ and $B_{jk}$:
\begin{equation}
    \begin{split}
        K_{jk}(x) &= A_{jk}\cos\left(\frac{\lambda_k}{\sqrt{D_j}}(x - \tilde{x}_{j - 1})\right) \\
        &+ B_{jk}\sin\left(\frac{\lambda_k}{\sqrt{D_j}}(x - \tilde{x}_{j - 1})\right)\,.
    \end{split}
\end{equation}
The solution $c(x, t)$ is subject to boundary, continuity and flux continuity conditions at interfaces $x_j$ as $j = 1, 2, \dots, m - 1$. This implies the following conditions,
\begin{align}
    \label{eq:analytical_solution_continuity}
    \lim_{x \rightarrow \tilde{x}_j^-} c(x, t) &= \lim_{x \rightarrow \tilde{x}_j^+} c(x, t)\,,\\
    \label{eq:analytical_solution_flux_continuity}
    \lim_{x \rightarrow \tilde{x}_j^-} D(x)\frac{\partial c}{\partial x} &= \lim_{x \rightarrow \tilde{x}_j^+} D(x)\frac{\partial c}{\partial x}\,,\\
   \label{eq:analytical_solution_boundary_conditions}
    c(\tilde{x}_0, t) = 0 \quad &\text{and} \quad c(\tilde{x}_m, t) = 0\,.
\end{align}
Substituting $K_j(x)$ and $T(x)$ into these equations yields,
\begin{align}
    \label{eq:analytical_solution_left_boundary_explicit}
    A_{1k} &= 0\,,\\
    \label{eq:analytical_solution_continuity_explicit}
    \begin{split}
     A_{j+1,k} &= A_{jk}\cos\left(\frac{\lambda_k (\tilde{x}_j - \tilde{x}_{j - 1})}{\sqrt{D_j}}\right) \\
        & + B_{jk}\sin\left(\frac{\lambda_k (\tilde{x}_j - \tilde{x}_{j - 1})}{\sqrt{D_j}}\right)
    \end{split}\\
    \label{eq:analytical_solution_flux_continuity_explicit}
    \begin{split}
    \frac{\sqrt{D_{j + 1}}}{\sqrt{D_j}}B_{j + 1, k}= & -A_{jk}\sin\left(\frac{\lambda_k(\tilde{x}_j - \tilde{x}_{j - 1})}{\sqrt{D_j}}\right) \\
        &+ B_{jk}\cos\left(\frac{\lambda_k(\tilde{x}_j - \tilde{x}_{j - 1})}{\sqrt{D_j}}\right)\,,
    \end{split}
\end{align}
and
\begin{align}
    \label{eq:analytical_solution_right_boundary_explicit}
        A_{mk}\cos\left(\frac{\lambda_k(\tilde{x}_m - \tilde{x}_{m - 1})}{\sqrt{D_m}}\right) 
        &+ B_{mk}\sin\left(\frac{\lambda_k(\tilde{x}_m - \tilde{x}_{m - 1})}{\sqrt{D_m}}\right)\nonumber\\ &= 0\,.
\end{align}
One can see that eqns. \eqref{eq:analytical_solution_left_boundary_explicit}, \eqref{eq:analytical_solution_continuity_explicit}, and \eqref{eq:analytical_solution_flux_continuity_explicit} represent a recurrence relation. One starts with $A_{1k} = 0$ and an unknown $B_{1k}$ to calculate $A_{jk}$ and $B_{jk}$ for $j = 2, 3, \dots, m$. These equations together with \eqref{eq:analytical_solution_right_boundary_explicit} can be converted into a form that is independent of $B_{1k}$. To this end, we introduce reduced coefficients $\hat{A}_{jk}$ and $\hat{B}_{jk}$ and a corresponding reduced spatial solution $\hat K_{jk}(x)$, such that
\begin{equation}
    \label{eq:A_jk_B_jk}
    A_{jk} = B_{1k}\hat{A}_{jk} \quad \text{and} \quad B_{jk} = B_{1k}\hat{B}_{jk}\,.
\end{equation}
Furthermore,
\begin{equation}
    \begin{split}
        \hat K_{jk}(x) &= \hat A_{jk}\cos\left(\frac{\lambda_k}{\sqrt{D_j}}(x - \tilde{x}_{j - 1})\right) \\
        &+ \hat B_{jk}\sin\left(\frac{\lambda_k}{\sqrt{D_j}}(x - \tilde{x}_{j - 1})\right)\,.
    \end{split}
\end{equation}
Notice that spatial solutions $K_{jk}(x)$ are now expressed in terms of unknown coefficients $B_{1k}$ and reduced spatial solutions $\hat K_{jk}(x)$, i.e.,
\begin{equation}
    \label{eq:K_jk_K_hat_jk}
    K_{jk}(x) = B_{1k}\hat K_{jk}(x)\,.
\end{equation}
This leads to a new set of equations
\begin{align}
    \label{eq:A_hat_1k_B_hat_1k}
    \hat{A}_{1k} =& 0 \quad \text{and} \quad \hat{B}_{1k} = 1\,,\\
    \label{eq:A_hat_jk}
    \begin{split}
        \hat{A}_{jk} =& \hat{A}_{j - 1, k}\cos\left(\frac{\lambda_k(\tilde{x}_{j - 1} - \tilde{x}_{j - 2})}{\sqrt{D_{j - 1}}}\right) \\
        &+ \hat{B}_{j - 1, k}\sin\left(\frac{\lambda_k(\tilde{x}_{j - 1} - \tilde{x}_{j - 2})}{\sqrt{D_{j - 1}}}\right)\,,
    \end{split}\\
    \label{eq:B_hat_jk}
    \begin{split}
        \hat{B}_{jk} =&
        -\sqrt{\frac{D_{j - 1}}{D_j}}\hat{A}_{j - 1, k}\sin\left(\frac{\lambda_k(\tilde{x}_{j - 1} - \tilde{x}_{j - 2})}{\sqrt{D_{j - 1}}}\right) \\
        &+ \sqrt{\frac{D_{j - 1}}{D_j}}\hat{B}_{j - 1, k}\cos\left(\frac{\lambda_k(\tilde{x}_{j - 1} - \tilde{x}_{j - 2})}{\sqrt{D_{j - 1}}}\right)\,,
    \end{split}\\
    \label{eq:A_hat_4k_B_hat_4k}
    \begin{split}
     0=  &\hat{A}_{mk}\cos\left(\frac{\lambda_k(\tilde{x}_m - \tilde{x}_{m - 1})}{\sqrt{D_m}}\right) \\
        &+ \hat{B}_{mk}\sin\left(\frac{\lambda_k(\tilde{x}_m - \tilde{x}_{m - 1})}{\sqrt{D_m}}\right)\,.
    \end{split}
\end{align}
The recurrence relations governed by equations \eqref{eq:A_hat_1k_B_hat_1k}, \eqref{eq:A_hat_jk} and \eqref{eq:B_hat_jk} together with \eqref{eq:A_hat_4k_B_hat_4k} represent a non-linear recursive equation in terms of $\lambda_k$. It is solved numerically using, e.g., a bracketing algorithm, since all roots of this equation are at zero-crossings. This results in multiple separation constants $\lambda_k$.

After solving the non-linear recursive equation above for $\lambda_k$, equations \eqref{eq:A_hat_1k_B_hat_1k}, \eqref{eq:A_hat_jk}, \eqref{eq:B_hat_jk} are reused to calculate the reduced coefficients $\hat{A}_{jk}$ and $\hat{B}_{jk}$ and the reduced spatial solutions $\hat K_{jk}(x)$. Then, according to \eqref{eq:K_jk_K_hat_jk} the spatial solutions $K_{jk}(x)$ can be written down in terms of the reduced spatial solutions $\hat K_{jk}(x)$ and the coefficients $B_{1k}$ which are essentially the expansion coefficients of the infinite series representing the solution to the original problem. To obtain the expansion coefficients $B_{1k}$, the following integration is carried out:
\begin{align}
    B_{1k} = \frac{1}{G_k} \sum_{j = 1}^m \int_{\tilde x_{j - 1}}^{\tilde x_j} \left(c_0(x) - c_s(x)\right)\hat K_{jk}(x)dx\,,
    \label{eq:B1k_integral}
\end{align}
where $G_k$ are normalization constants. The normalization constants are found by the following integration:
\begin{align}
    G_k = \sum_{j = 1}^m \int_{\tilde x_{j - 1}}^{\tilde x_j} \hat K_{jk}^2(x)dx\,,
    \label{eq:Gk_integral}
\end{align}
The closed-form formulas for the integrals \eqref{eq:B1k_integral} and \eqref{eq:Gk_integral} are presented in Appendix \ref{app:integrals}, see eqns. \eqref{eq:app_B1k} and \eqref{eq:app_Gk}, respectively.

\begin{figure*}
    \centering
    \includegraphics[width=.95\textwidth]{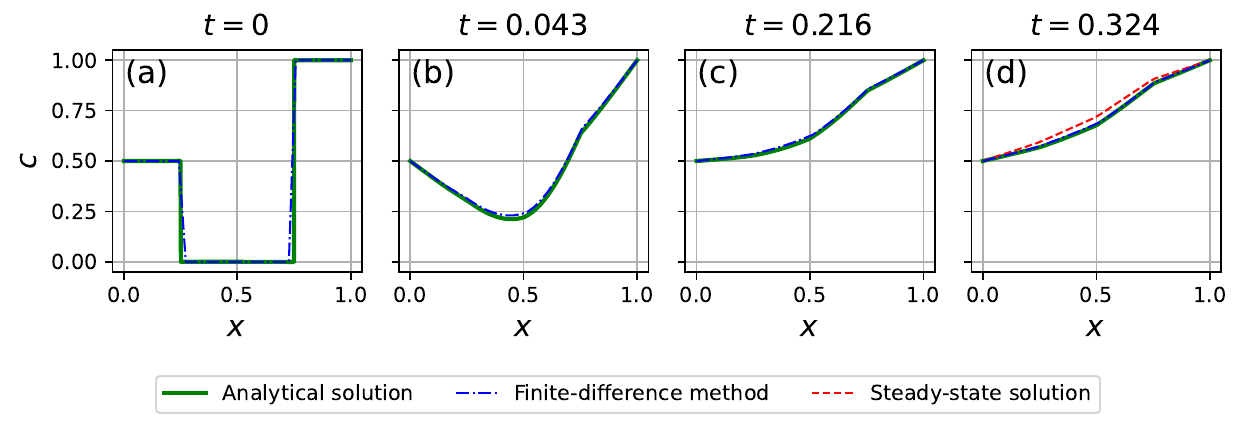}
    \caption{Analytical solution compared at different time instants together with the finite-difference solution. Panel (a) shows the initial conditions, panels (b) and (c) depict time evolution at different time instants, and panel (d) demonstrates how the analytical and numerical finite-difference solutions tend to the steady state solution.}
    \label{fig:ana_vs_fdm}
\end{figure*}

To compute the analytical solution one first solves the non-linear recursive equation governed by \eqref{eq:A_hat_1k_B_hat_1k}, \eqref{eq:A_hat_jk}, \eqref{eq:B_hat_jk} and \eqref{eq:A_hat_4k_B_hat_4k} to obtain the separation constants $\lambda_k$. Then the obtained $\lambda_k$ are used to compute the reduced coefficients $\hat{A}_{jk}$ and $\hat{B}_{jk}$ using the recurrence relation \eqref{eq:A_hat_1k_B_hat_1k}, \eqref{eq:A_hat_jk}, and \eqref{eq:B_hat_jk}. Subsequently, the reduced spatial solutions $\hat K_{jk}(x)$ are computed as well. Next, the normalization constants $G_k$ are computed. The expansion coefficients $B_{1k}$ are computed afterwards and combined with $\hat{A}_{jk}$ and $\hat{B}_{jk}$ using \eqref{eq:A_jk_B_jk} to produce the coefficients $A_{jk}$ and $B_{jk}$. Finally, the piecewise solution is written as
\begin{equation}
    \label{eq:analytical_solution}
        c(x, t) = c_s(x) + \sum_{k = 1}^\infty\exp\left({-\lambda_k^2 t}\right)
        K_{jk}(x)\,, 
\end{equation}
for $\tilde{x}_{j - 1} \leq x < \tilde{x}_j$. The graph of the analytical solution \eqref{eq:analytical_solution} is shown in Fig.~\ref{fig:ana_vs_fdm}. A system consisting of 4 layers with diffusion coefficients $D_1 = D_4 = 1$, $D_2 = 0.75$, and $D_3 = 0.5$ and initial conditions $c_{0,1} = 0.5$, $c_{0,2} = c_{0,3} = 0$, and $c_{0,4} = 1$ serves as an example. Fig.~\ref{fig:ana_vs_fdm}(a) corresponds to the initial conditions, Fig.~\ref{fig:ana_vs_fdm}(b) and Fig.~\ref{fig:ana_vs_fdm}(c) depict the time evolution of the analytical and finite-difference solutions, and Fig.~\ref{fig:ana_vs_fdm}(d) demonstrates that both the analytical and the finite-difference solutions tend to the steady state solution as time tends to infinity.

\subsection{Relaxation to steady state}
\begin{figure*}
    \centering
    \includegraphics[width=0.9\textwidth]{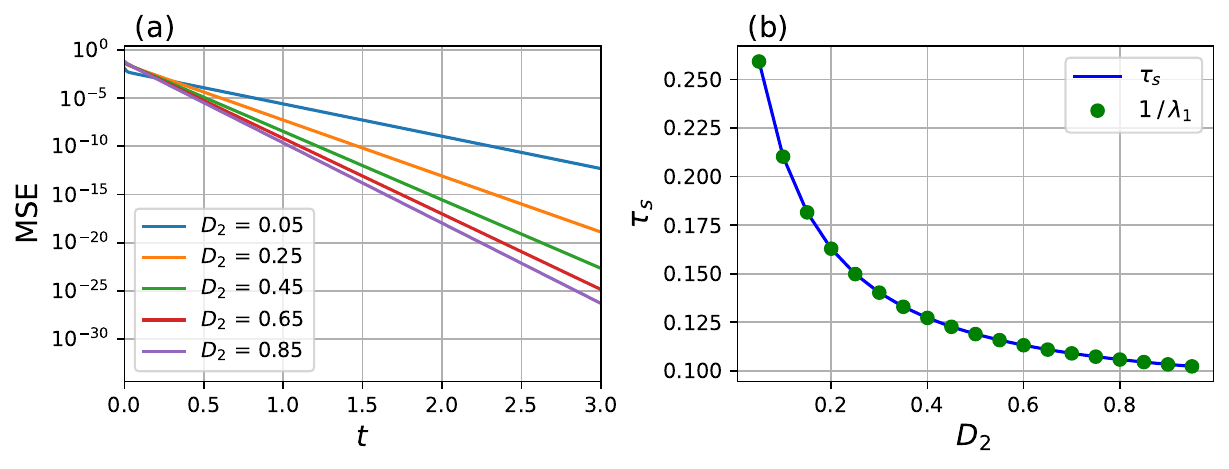}
    \caption{Relaxation rate analysis of the problem. (a) The function $\delta(t)$ for various values of $D_2$. (b) The time constant $\tau_s$ and smallest positive eigenvalue reciprocals for various values of $D_2$.}
    \label{fig:relaxation_rate_analysis}
\end{figure*}

The standard measure of the relaxation rate is in terms of the principal non-zero eigenvalue of the negative Laplacian \cite{bressloff2022}. Here, we obtain a similar result in terms of the smallest separation constant $\lambda_1$. To this end, the function $\delta(t)$ depicted in Fig.~\ref{fig:relaxation_rate_analysis}(a) is introduced to measure the agreement between the analytical solution at time $t$ and the steady state solution:
\begin{equation}
    \label{eq:delta_t}
    \delta(t) = \frac{1}{\tilde x_2 - \tilde x_0} \lVert c(x, t) - c_s(x) \rVert_2^2\,.
\end{equation}
Due to the analytical solution exhibiting exponential decay, the following asymptotic equivalence holds for $\delta (t)$:
\begin{equation}
    \label{eq:asymptotic_equivalence}
    \delta(t) \sim C\exp\left(-\frac{t}{\tau_s}\right) \quad \text{as} \quad t \rightarrow \infty\,,
\end{equation}
where $C$ is a non-zero constant and $\tau_s$ is the time constant. The time constant $\tau_s$, which is the measure of the relaxation rate, was calculated for different values of $D_2$ using data depicted in Fig.~\ref{fig:relaxation_rate_analysis}(a). In Fig.~\ref{fig:relaxation_rate_analysis}(b) we compare it with the reciprocal of the separation constant $\lambda_1$, which demonstrates that the time constant is indeed the smallest separation constant reciprocal of the analytical solution expansion, i.e., $\tau_s =1/ \lambda_1$. Hence, the relaxation rate can be expressed in terms of the smallest separation constant $\lambda_1$ and is inversely proportional to $D_2$.

\subsection{Concentration contrast and diffusivity ratio}
Due to the piecewise nature of the problem, the solution may exhibit non-smooth behavior, manifested as concentration kinks with pronounced gradients. This, in turn, can lead to the build-up of concentration contrasts that may hinder the operation of electrolyzers. To identify the conditions under which such situations arise, we introduce the diffusivity ratio $r_D$ as
\begin{equation}
    \label{eq:diffusivity_ration}
    r_D = \frac{D_1}{D_2}\,,
\end{equation}
and analyze the steady state solutions by varying $r_D$. Moreover, we vary the interface position $\tilde x_1$ and the boundary concentration $c_A$ to assess whether the influence of $r_D$ on the build-up of the concentration contrast is robust.

\begin{figure*}
    \centering
    \includegraphics[width=0.9\textwidth]{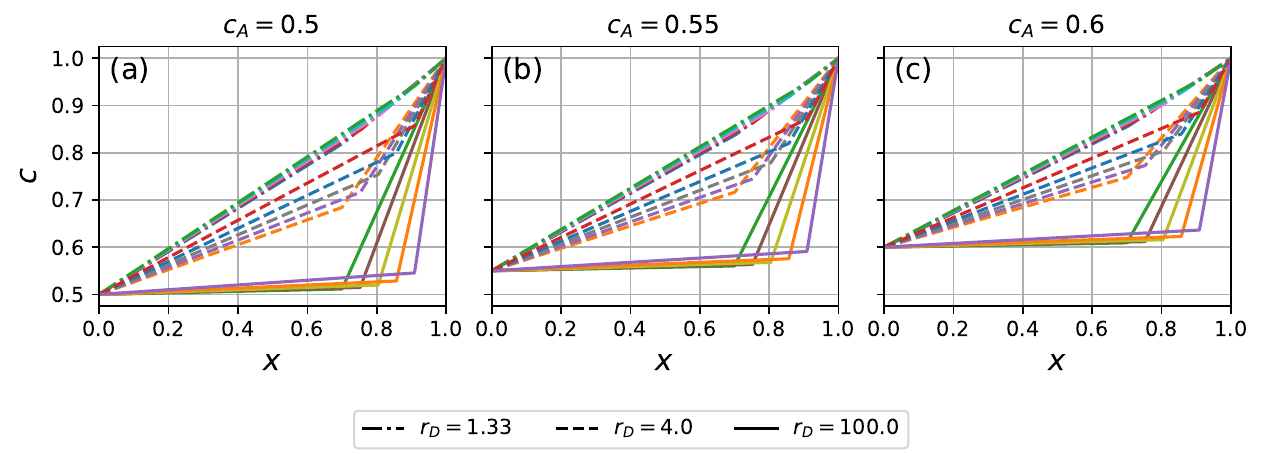}
    \caption{Steady state solutions compared for various values of diffusivity ratio $r_D$, boundary concentration $c_A$, and interface position $\tilde x_1$. Panels (a)--(c) correspond to $c_A = 0.5$, $0.55$ and $0.6$, respectively. Lines with the same style correspond to a fixed value of $r_D$ (see legend), whereas different colors indicate different interface positions $\tilde x_1 \in [0.7, 0.9]$.}
    \label{fig:steady_state_gradients}
\end{figure*}

Figure \ref{fig:steady_state_gradients} presents the steady state solutions, where $r_D$ was varied from 1.33 via 4.0 to 100, $\tilde x_1$ from $0.7$ to $0.9$ in steps of 0.05, and $c_A$ from $0.5$ to $0.6$ in steps of 0.05. The results show that perturbations in the boundary concentration and the interface position mainly affect the location of the kink-point. In contrast, the diffusivity ratio $r_D$ strongly influences the gradient to the left and right of the kink-point. Hence, the stationary solutions indicate that pronounced hydroxide concentration gradients occur only when the diffusivity ratio $r_D$ takes very large values than approximately 50.

\subsection{Finite difference numerical method}
The numerical solution of the diffusion equation \eqref{eq:diffusion_equation} can be obtained using the finite difference method (FDM). In the present case, we will compare our quantum algorithm also with a classical numerical integration scheme.  With respect to the quality of conservation, two types of FDM schemes can be distinguished, conservative and non-conservative. Although the non-conservative FDM may converge faster than the conservative FDM for smooth solutions, it fails to correctly capture discontinuities \cite{neelan2025physics}. Since discontinuities arise in Eq. \eqref{eq:diffusion_equation} due to discontinuous diffusivity profile, we employ conservative FDM. 

First, the problem is discretized uniformly in space and time, so that $x_j = x_0 + j\Delta x$ and $t_l = t_0 + l\Delta t$, with $x_0 = 0$ and $t_0 = 0$. The indices $j$ and $l$ satisfy $j = 0, \dots, N + 1$ and $l = 0, \dots, M$. The spatial interval $x \in [\tilde x_0, \tilde x_m]$ is divided into $N$ internal and $2$ boundary nodes; therefore, $\Delta x = (\tilde x_m - \tilde x_0) \mathbin{/} (N + 1)$. To obtain the conservative scheme, the time derivative is discretized by a forward finite difference, while the outer and inner spatial derivatives are discretized by a central finite difference. The fluxes are evaluated at the staggered points $x_{j - \frac{1}{2}}$ and $x_{j + \frac{1}{2}}$, and subsequently approximated using concentrations at the grid points:
\begin{equation}
    \begin{split}
        \frac{c_j^{l + 1} - c_j^l}{\Delta t} = \frac{1}{\Delta x^2}\Big[&D_{j + \frac{1}{2}}\left(c_{j + 1}^l - c_j^l\right) -\\
        & D_{j - \frac{1}{2}}\left(c_j^l - c_{j - 1}^l\right)\Big]\,.
    \end{split}
\end{equation}
Defining
\begin{equation}
    r_j = \frac{D_j \Delta t}{\Delta x^2} \quad \text{and} \quad s_j = 1 - \frac{D_{j + \frac{1}{2}} \Delta t}{\Delta x^2} - \frac{D_{j - \frac{1}{2}} \Delta t}{\Delta x^2}
\end{equation}
allows for a more explicit expression:
\begin{equation}
    \begin{split}
        c_j^{l + 1} &= r_{j + \frac{1}{2}} c_{j + 1}^l + s_j c_j^l + r_{j - \frac{1}{2}} c_{j - 1}^l\,.
    \end{split}
\end{equation}
The scheme can be expressed as a system of linear equations:
\begin{equation}
    c^{l + 1} = Ac^l\,,
\end{equation}
where matrix $A \in \mathbb{R}^{(N + 2) \times (N + 2)}$ has the following structure
\begin{equation}
    A = \begin{bmatrix}
        1 & 0 & 0 & 0 & \dots & 0 & 0 & 0 \\
        r_\frac{1}{2} & s_1 & r_\frac{3}{2} & 0 & \dots & 0 & 0 & 0 \\
        0 & r_\frac{3}{2} & s_2 & r_\frac{5}{2} & \dots & 0 & 0 & 0 \\
        \vdots & \vdots & \ddots & \ddots & \ddots & \vdots & \vdots & \vdots \\
        0 & 0 & 0 & 0 & \dots & r_{N - \frac{1}{2}} & s_{N} & r_{N + \frac{1}{2}} \\
        0 & 0 & 0 & 0 & \dots & 0 & 0 & 1 \\
    \end{bmatrix}\,.
\end{equation}
The first and the last rows of $A$ serve to incorporate the inhomogeneous Dirichlet boundary conditions located at $c_0^l$ and $c_{N + 1}^l$. To ensure that this scheme is stable, the generalization of the Courant–Friedrichs–Lewy (CFL) condition, namely, the principle of frozen coefficients is used, i.e.:
\begin{equation}
    \Delta t < \frac{\Delta x^2}{2 \max_x D(x)}\,.
\end{equation}
It was proven in ref. \cite{lee2017stability} that the CFL condition is sufficient for the conservative scheme applied to the diffusion equation with discontinuous coefficients.

\section{Variational quantum algorithm}

\subsection{Weak formulation framework including boundary conditions}
\begin{figure*}
    \centering
    \begin{minipage}[c]{0.54\textwidth}
        \includegraphics[width=\textwidth]{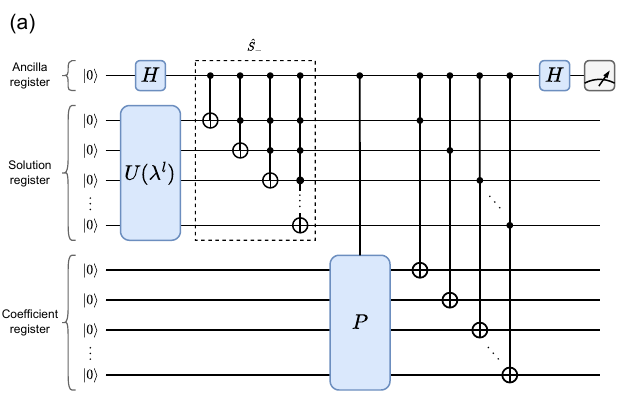}
    \end{minipage}
    \hfill
    \begin{minipage}[c]{0.43\textwidth}
        \includegraphics[width=\textwidth]{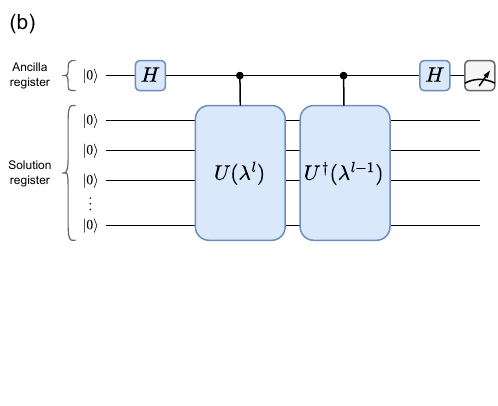}
    \end{minipage}
    \caption{Quantum circuits for computing individual terms of the cost function. Circuit (a) evaluates the terms $S_\text{PER}$ and $S^\pm$, where the unitary $P$ incorporates the respective coefficients. Circuit (b) evaluates the term $S_\text{LIN}$.}
    \label{fig:quantum_circuits}
\end{figure*}

We apply the weak formulation framework proposed by Bengoechea et al. \cite{bengoechea2025towards}. In a weak formulation, instead of requiring the differential equation to hold pointwise, it is tested against functions $v \in V$, where $V$ is, in the present case, a Sobolev space. This results in linear operator equations in integral form, the weak formulation of the original equation. For the diffusion problem considered here, the linear operator equations are equivalent to the optimality condition of a minimization of a quadratic functional over the function space $V$. Replacing the originally infinite-dimensional $V$ with a finite-dimensional subspace then yields a spatial discretization of the problem and produces a finite-dimensional quadratic minimization problem amenable to the VQA paradigm. For the approach to be applicable to \eqref{eq:diffusion_equation}, the problem must be slightly reformulated by recognizing that the charge profile $c(x, t)$ can be decomposed into two components, namely, the steady state solution and the transient solution:
\begin{equation}
    \label{eq:ss_ts_decomposition}
    c(x, t) = c_s(x) + \tilde c(x,t) \,,
\end{equation}
where $\tilde c(x, t)$ is the transient solution. Then, eq. \eqref{eq:diffusion_equation} is reformulated in terms of the transient solution $\tilde c(x, t)$:
\begin{equation}
    \label{eq:diffusion_equation_transient}
    \frac{\partial \tilde c}{\partial t} = \frac{\partial}{\partial x}\left(D(x)\frac{\partial \tilde c}{\partial x}\right) \quad \text{in} \quad \Omega_T = (\tilde{x}_0, \tilde{x}_m) \times (0, T]
\end{equation}
with the following initial,
\begin{equation}
    \tilde c(x, 0) = c_0(x) - c_s(x)\,,
\end{equation}
and homogeneous Dirichlet boundary conditions,
\begin{equation}
    \tilde c(\tilde{x}_0, t) = 0 \quad \mbox{and} \quad \tilde c(\tilde{x}_m, t) = 0\,.
\end{equation}
Equation \eqref{eq:diffusion_equation_transient}, expressed in terms of the transient solution $\tilde c(x, t)$, is now ready to be reformulated as a finite-dimensional optimization problem. The time derivative is approximated using the backward finite difference formula:
\begin{equation}
    \frac{\partial \tilde c}{\partial t} = \frac{\tilde c^l - \tilde c^{l - 1}}{\Delta t} + \mathcal{O}(\Delta t)
\end{equation}
where $\tilde c^l = \tilde c^l(x)$ is the spatial charge profile at $l$-th time instant, and $\Delta t$ is the integration time step. Therefore, the PDE is approximated by a second-order linear system of ODEs:
\begin{equation}
    \label{eq:semi_discrete_diffusion_equation}
    \tilde c^l - \Delta t\frac{d}{dx}\left(D(x)\frac{d\tilde c^l}{dx}\right) = \tilde{c}^{l - 1}\,.
\end{equation}
The weak formulation of the obtained system of ODEs is given by
\begin{equation}
    \label{eq:weak_formulation}
    \underbrace{\int_\Omega v^l\left[\frac{\tilde c^l}{\Delta t} - \frac{d}{dx}\left(D(x)\frac{d\tilde c^l}{dx}\right)\right]dx}_{a(v^l, \tilde c^l)} = \underbrace{\int_\Omega v^l\frac{\tilde c^{l - 1}}{\Delta t}dx}_{f(v^l)}\,,
\end{equation}
for all test functions $v^l\in V$. Here, $a(v^l, \tilde c^l)$ is a symmetric and positive definite bilinear form and $f(v^l)$ is a linear form. The weak formulation above is equivalent to the following optimization problem,
\begin{equation}
    \label{eq:weak_formulation_optimization}
    \min_{\tilde c^l \in V} \left[\frac{1}{2}a(\tilde c^l, \tilde c^l) - f(\tilde c^l)\right]\,.
\end{equation}
According to the Ritz-Galerkin method, the space $V$ is replaced with a finite-dimensional space $\mathbb{R}^N$, leading to a finite-dimensional optimization problem. Approximating integrals in $a(\tilde c^l, \tilde c^l)$ and $f(\tilde c^l)$ using the midpoint rule results in:
\begin{align}
    \label{eq:discrete_weak_formulation_optimization}
    \begin{split}
        \min_{\tilde c^l \in \mathbb{R}^N} \Bigg\{&\frac{\Delta x}{2 \Delta t}\sum_{j = 1}^N [\tilde c^l]^2
         - \frac{\Delta x}{2}\sum_{j = 1}^N \tilde c^l_j\left[\frac{d}{dx}\left(D(x)\frac{d\tilde c^l}{dx}\right)\right]\bigg\rvert_{x = x_j} \\
        & - \frac{\Delta x}{\Delta t}\underbrace{\sum_{j = 1}^N\tilde c^{l - 1}_j\tilde c^l_j}_{S_{\text{LIN}}(\tilde c^l)}\Bigg\}\,.
    \end{split}
\end{align}
The diffusion operator in \eqref{eq:discrete_weak_formulation_optimization} is discretized by successive application of the central finite difference formula using staggered (outer derivative) and collocated (inner derivatives) points, see \eqref{eq:diffusion_operator_discretization} in Appendix \ref{app:vqa}. Then the second sum in \eqref{eq:discrete_weak_formulation_optimization} becomes:
\begin{equation}
    \begin{split}
         \sum_{j = 1}^N \tilde c^l_j&\left[\frac{d}{dx}\left(D(x)\frac{d\tilde c^l}{dx}\right)\right]\bigg\rvert_{x = x_j} 
        = \frac{1}{\Delta x^2}\sum_{j = 1}^N \tilde c^l_jD_{j + \frac{1}{2}}\tilde c^l_{j + 1} \\
        & - \frac{1}{\Delta x^2}\underbrace{\sum_{j = 1}^N\tilde c_j^l(D_{j - \frac{1}{2}} + D_{j + \frac{1}{2}})\tilde c_j^l}_{S^{\pm}(c^l)} \\
        & + \frac{1}{\Delta x^2}\sum_{j = 1}^N \tilde c^l_jD_{j - \frac{1}{2}}\tilde c^l_{j - 1}\,.
    \end{split}
    \label{eq:vqa_diffusion_operator_discretized}
\end{equation}
Since the transient solution $\tilde c(x, t)$ is subject to homogeneous Dirichlet conditions $\tilde c_0 = \tilde c_{N + 1} = 0$, the first and the last sums in the equation above can be combined as (see \eqref{eq:app_S_DIR} in Appendix \ref{app:vqa} for detailed derivation):
\begin{equation}
    \begin{split}
         S_{\text{DIR}}(\tilde c^l) = 2\sum_{j = 1}^{N - 1} \tilde c^l_jD_{j + \frac{1}{2}}\tilde c^l_{j + 1}\,.
    \end{split}
    \label{eq:S_DIR}
\end{equation}
The desired Dirichlet sum $S_\text{DIR}$ is challenging to evaluate directly on a quantum computer due to its incorporation of boundary conditions that break translational symmetry. Bengoechea et al. \cite{bengoechea2025towards} propose decomposing this sum into a periodic one,
\begin{align*}
    S_{\text{PER}}(\tilde c^l) = \tilde c_1D_{N + \frac{1}{2}}\tilde c_N + \sum_{j = 1}^{N - 1} \tilde c^l_jD_{j + \frac{1}{2}}\tilde c^l_{j + 1}\,,
\end{align*}
and an additional corrective term, which can be evaluated using the circuit proposed by Over et al. \cite{over2025boundary}. Therefore, the sum $S_{\text{DIR}}(\tilde c^l)$ is expressed in terms of the periodic sum $S_{\text{PER}}(\tilde c^l)$ and the corrective term $S_{\text{BND}}(\tilde c^l)$:
\begin{align*}
        S_{\text{DIR}}(\tilde c^l) &= 2S_{\text{PER}}(\tilde c^l) - \underbrace{2\tilde c_1D_{N + \frac{1}{2}}\tilde c_N}_{S_{\text{BND}}(\tilde c^l)} \,.
\end{align*}
Then, the sum involving the diffusion operator evaluated at nodes becomes
\begin{align*}
    & \sum_{j = 1}^N \tilde c^l_j\left[\frac{d}{dx}\left(D(x)\frac{d\tilde c^l}{dx}\right)\right]\bigg\rvert_{x = x_j} \\
    & = \frac{1}{\Delta x^2}S_{\text{DIR}}(\tilde c^l) - \frac{1}{\Delta x^2}S^{\pm}(c^l) \\
    & = \frac{2}{\Delta x^2}\left[S_{\text{PER}}(\tilde c^l) - \frac{1}{2}S_{\text{BND}}(\tilde c^l)\right] - \frac{1}{\Delta x^2}S^{\pm}(c^l)\,.
\end{align*}
Substituting this result into eq. \eqref{eq:discrete_weak_formulation_optimization} gives the optimization problem with a cost function that consists of terms that can be evaluated directly using quantum circuits:
\begin{align}
    \label{eq:discrete_weak_formulation_optimization_2}
    \begin{split}
        \min_{\tilde c^l \in \mathbb{R}^N} \Bigg\{&\frac{\Delta x}{2 \Delta t}\sum_{j = 1}^N [\tilde c^l]^2 - \frac{1}{\Delta x}S_{\text{PER}}(\tilde c^l) + \frac{1}{2\Delta x}S_{\text{BND}}(\tilde c^l) \\
        &+ \frac{1}{2\Delta x}S^\pm(\tilde c^l) - \frac{\Delta x}{\Delta t}S_{\text{LIN}}(\tilde c^l)\Bigg\}\,.
    \end{split}
\end{align}
Since the cost function is supposed to be expressed as a function of the normalized charge profile $\hat{c}^l \in \mathbb{R}^N$ the following substitution is made:
\begin{align}
    \label{eq:lambda_0}
    \tilde c_j^l = \lambda_0^l\hat{c}_j^l\,, & \quad j = 1, \dots, N\,,\end{align}
where $\lambda_0^l$ is the normalization constant and $\|\hat{c}^l\|_2=1$. Finally, the transformed version of the optimization problem \eqref{eq:discrete_weak_formulation_optimization} suitable for an implementation as a VQA algorithm is given by
\begin{align}
    \label{eq:discrete_weak_formulation_optimization_3}
    \begin{split}
        \min_{\lambda_0^l \in \mathbb{R}, \space \hat{c}^l \in \mathbb{R}^N} \Bigg\{&\frac{\Delta x}{2\Delta t}[\lambda_0^l]^2 - \frac{1}{\Delta x}[\lambda_0^l]^2S_{\text{PER}}(\hat{c}^l) \\
        &+ \frac{1}{2\Delta x}[\lambda_0^l]^2S_{\text{BND}}(\hat{c}^l) + \frac{1}{2\Delta x}[\lambda_0^l]^2S^\pm(\hat{c}^l) \\
        &- \frac{\Delta x}{\Delta t}\lambda_0^{l - 1}\lambda_0^lS_{\text{LIN}}(\hat{c}^l)\Bigg\}\,.
    \end{split}
\end{align}
The normalized charge profile $\hat c^l$ is generated by a parameterized quantum circuit, that is, it depends on the parameter vector $\bm \lambda$
\begin{equation}
    \label{eq:c_hat_l}
    \hat c^l = \hat c(\bm \lambda^l) = U(\bm \lambda^l)|0\rangle^{\otimes n}\,.
\end{equation}
Each term $S_\text{PER}$, $S^\pm$, $S_\text{LIN}$ and $S_\text{BND}$ is computed using a corresponding Hadamard test circuit, see also the appendix of ref. \cite{Ingelmann2024}. The circuit diagrams for computing $S_\text{PER}$, $S^\pm$ and $S_\text{LIN}$ are displayed in Fig.~\ref{fig:quantum_circuits}. The quantum circuit for $S_\text{BND}$ is introduced in \cite{over2025boundary}.
The first register containing a single qubit is an ancillary qubit of the Hadamard test. The solution register allows loading the amplitude encoding of the charge concentrations and performing operations on them. The purpose of the coefficient register is to introduce the coefficients of the quadratic forms $S_\text{PER}$ and $S^\pm$. There, the $P$ unitary is nothing else but state preparation circuits that prepare states proportional to $(D_\frac{3}{2}, D_\frac{5}{2}, \dots, D_{N + \frac{1}{2}})$ for $S_\text{PER}$ and $(D_\frac{1}{2} + D_\frac{3}{2}, D_\frac{3}{2} + D_\frac{5}{2}, \dots, D_{N - \frac{1}{2}} + D_{N + \frac{1}{2}})$ for $S^\pm$ respectively. When computing $S^\pm$, the shift block $\hat S_-$ is omitted. The next section will introduce the quantum state preparation algorithm for such states.

\subsection{Quantum state preparation algorithm}
In this section, we present the quantum state preparation algorithm designed to encode the coefficients of quadratic forms in the VQA cost function. The necessity of this specific state preparation algorithm arises from the structure of these coefficient vectors, which predominantly consist of a small number of constant subvectors, a property directly resulting from the piecewise-constant nature of the diffusion coefficient. By exploiting this structure, our algorithm generates quantum circuits with significantly fewer gates compared to general-purpose state preparation methods.

\subsubsection{Successive bisection}
The underlying principle of the algorithm employs a recursive dichotomy process: starting with an initial constant vector, the procedure iteratively bisects it into constant subsegments using a finite number of iterations until the constructed vector precisely matches the desired target state. In each iteration, the process ensures that the approximation resulting from the bisection can be progressively refined in subsequent iterations. This is achieved by guaranteeing that when a segment is bisected, its magnitude is redistributed among its constant subsegments, such that they approximate the desired vector in an average sense. At the end of the process, the approximated vector is equal to the desired vector element-wise.
\begin{figure*}
    \centering
    \includegraphics[width=.9\textwidth]{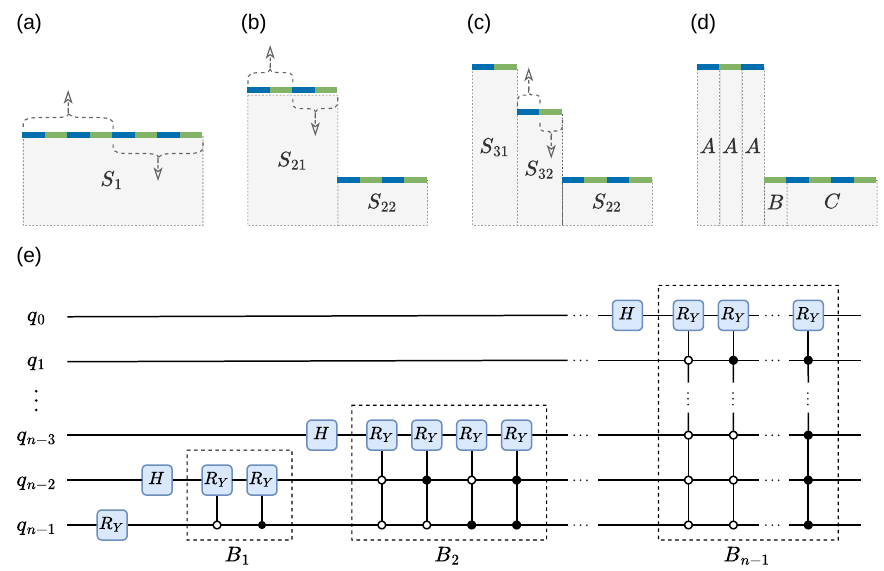}
    \caption{Sketch of the principle of the state preparation algorithm shown in panels (a)--(d) and a general quantum circuit produced by the state preparation algorithm shown in panel (e). The arrows in panels (a)--(d) indicate the direction in which each corresponding segment is shifted — either upward or downward. Gray rectangles with dashed borders depict the areas under the various segments. In panels (e), blocks $B_1$ and $B_2$ produce the second and the third approximations of the desired state, respectively. Block $B_{n - 1}$ produces the final approximation.}
    \label{fig:state_preparation_principle_and_circuit}
\end{figure*}

Figure~\ref{fig:state_preparation_principle_and_circuit}(a)--(d) provides an example of an 8-dimensional vector illustrating this principle. The desired vector is shown in Fig.~\ref{fig:state_preparation_principle_and_circuit}(d), it consists of a 3-element constant segment followed by a 5-element constant segment. To arrive at the desired vector, one starts with a constant vector as in Fig.~\ref{fig:state_preparation_principle_and_circuit}(a). Its magnitude is set equal to the combined magnitude of the target vector, i.e., $S_1 = 3A + B + C$. To approximate the desired vector, the initial constant vector is bisected into two segments $S_{21}$ and $S_{22}$, which are separated by moving $S_{21}$ upward and $S_{22}$ downward. This produces the first approximation of the desired vector shown in Fig.~\ref{fig:state_preparation_principle_and_circuit}(b). As can be seen, the magnitude of the left segment is $S_{21} = 3A + B$ and the magnitude of the right segment is $S_{22} = C$. The first approximation is then compared with the desired vector to determine which segments of it have to be further bisected. For example, bisecting $S_{22}$ is not necessary, because its counterpart in the desired vector is already constant, and there is no need to redistribute the magnitude of $S_{22}$. The left segment $S_{21}$ in the first approximation, however, has to be bisected, because it should eventually exhibit a discontinuity as in the desired vector. Thus, bisecting and separating $S_{21}$ produces the second approximation shown in Fig.~\ref{fig:state_preparation_principle_and_circuit}(c), where the magnitude $S_{21}$ is redistributed between the subsegments with magnitudes $S_{31} = 2A$ and $S_{32} = A + B$, respectively. Now, the subsegments $S_{31}$ and $S_{32}$ are compared with their counterparts in the desired vector. The subsegment $S_{31}$ does not have to be bisected, as its counterpart in the desired vector is already constant, thus it is left unchanged. The subsegment $S_{32}$, on the other hand, has to be bisected, as its counterpart in the desired vector has a discontinuity. Since $S_{32}$ is a two-element subsegment, it is bisected into two one-element subsegments with magnitudes $A$ and $B$, or, equivalently, two individual amplitudes with values $\sqrt A$ and $\sqrt B$. Therefore, bisecting $S_{32}$ produces the third and final approximation, which reproduces the desired vector exactly.

The example described above can be generalized into the following procedure: for each marked constant segment in the current approximation, (1) check if its counterpart in the desired vector contains a discontinuity; if so, (2) bisect the segment into two smaller constant subsegments ensuring correct magnitudes, and (3) mark the subsegments for the next iteration. 

We now formalize this procedure using the quantum state notation. Suppose a desired real-valued quantum state $\left|\psi\right\rangle \in \mathbb{C}^N$ containing normalized coefficients of a quadratic form is given. If it contains a discontinuity, which is the case for the given state, it is split into two partial states $|\tilde\psi_0\rangle$ and $|\tilde\psi_1\rangle$ with respect to the most significant qubit, i.e.,
\begin{equation}
    \label{eq:psi_splitting}
    |\psi\rangle = |0\rangle \otimes |\tilde\psi_0\rangle + |1\rangle \otimes |\tilde\psi_1\rangle \,.
\end{equation}
Then, the state $|\phi\rangle$, which is the first approximation of $|\psi\rangle$, is introduced and initialized such that
\begin{equation}
    |\phi\rangle \longleftarrow \sqrt{\langle \tilde\psi_0 | \tilde\psi_0 \rangle}|0\rangle \otimes |\tilde\phi_0\rangle + \sqrt{\langle \tilde\psi_1 | \tilde\psi_1 \rangle}|1\rangle \otimes |\tilde\phi_1\rangle \,,
\end{equation}
with
\begin{equation}
    |\tilde\phi_0\rangle \longleftarrow |+\rangle^{\otimes (n - 1)} \quad \text{and} \quad |\tilde\phi_1\rangle \longleftarrow |+\rangle^{\otimes (n - 1)} \,.
\end{equation}
Therefore, in the first iteration, $|\psi\rangle$ is approximated by $|\phi\rangle$, which consists of two uniform superpositions, such that their norms are equal to those of $|\tilde\psi_0\rangle$ and $|\tilde\psi_1\rangle$. Figure~\ref{fig:state_preparation_principle_and_circuit}(e) with blocks $B_1$, $B_2$ and $B_{n - 1}$ being omitted, shows the circuit for preparing the state $|\phi\rangle$. The amplitudes of the partial states $|\tilde\phi_0\rangle$ and $|\tilde\phi_1\rangle$ in $|\phi\rangle$ are assigned using a rotation about the $y$-axis ($R_Y$) acting on the most significant qubit $q_{n - 1}$, while partial states $|\tilde\phi_0\rangle$ and $|\tilde\phi_1\rangle$ themselves are generated using the Hadamard gates acting on qubits $q_0, q_1, \dots, q_{n - 2}$. Application of the $R_Y$ gate can be thought of as redistributing the unit norm of the initial uniform superposition between two constant segments as in Figs~\ref{fig:state_preparation_principle_and_circuit}(a)-(b). Pairs of states $(|\tilde\psi_0\rangle, |\tilde\phi_0\rangle)$ and $(|\tilde\psi_1\rangle, |\tilde\phi_1\rangle)$ are then marked as the ones requiring further processing. In the second iteration, the marked states $|\tilde\psi_0\rangle$ and $|\tilde\psi_1\rangle$ are checked for discontinuities. If among the marked states there are ones containing discontinuities, the algorithm proceeds further by splitting these states, denoted as $|\tilde\psi_{j}\rangle$ (where $j \in \{0, 1\}$), analogously to eq. \eqref{eq:psi_splitting},
\begin{equation}
    \label{eq:psi_splitting_2}
    |\tilde\psi_j\rangle = |0\rangle \otimes |\tilde\psi_{j0}\rangle + |1\rangle \otimes |\tilde\psi_{j1}\rangle \,.
\end{equation}
and the state $|\phi\rangle$ is updated by bisecting the partial state $|\tilde\phi_j\rangle$ as,
\begin{equation*}
    |\tilde\phi_j\rangle \longleftarrow \sqrt{\frac{\langle \tilde\psi_{j0} | \tilde\psi_{j0} \rangle}{\langle \tilde\psi_j |\tilde\psi_j \rangle}}|0\rangle \otimes |\tilde\phi_{j0}\rangle + \sqrt{\frac{\langle \tilde\psi_{j1} | \tilde\psi_{j1} \rangle}{\langle \tilde\psi_j | \tilde\psi_j \rangle}}|1\rangle \otimes |\tilde\phi_{j1}\rangle \,,
\end{equation*}
with
\begin{equation}
    |\tilde\phi_{j0}\rangle \longleftarrow |+\rangle^{\otimes (n - 2)} \quad \text{and} \quad |\tilde\phi_{j1}\rangle \longleftarrow |+\rangle^{\otimes (n - 2)} \,,
\end{equation}
whereas pairs of states $(|\tilde\psi_{j0}\rangle, |\tilde\phi_{j0}\rangle)$ and $(|\tilde\psi_{j1}\rangle, |\tilde\phi_{j1}\rangle)$ are marked for the next iteration. For example, if the state in Fig.~\ref{fig:state_preparation_principle_and_circuit}(d) is to be generated, the partial state $|\tilde\phi_0\rangle$ is bisected in the second iteration, as it corresponds to the left subsegment $S_{21}$. The updated state $|\phi\rangle$ is generated by introducing the block $B_1$ in the circuit in Figure~\ref{fig:state_preparation_principle_and_circuit}(e). Depending on which partial states $|\tilde\phi_j\rangle$ have to be bisected, corresponding controlled $R_Y$ (CRY) gates are introduced, such that they act on qubit $q_{n - 2}$ and controlled by qubit $q_{n - 1}$, while conditioned by the computational state $|j\rangle$. Similarly, in the third iteration, partial states $|\tilde\psi_{j0}\rangle$ and $|\tilde\psi_{j1}\rangle$ are checked for discontinuities, and for states $|\tilde\psi_{jk}\rangle$ (with $(j, k) \in \{0, 1\}^2$) containing discontinuities, corresponding states $|\tilde\phi_{jk}\rangle$ are bisected. However, this time the block $B_2$ is introduced, where multi-controlled $R_Y$ (MCRY) gates are conditioned by the state $|jk\rangle$. Therefore, starting from the third iteration, the process is similar except that MCRY gates are used to carry out the bisection process. The complete algorithm is presented in Algorithm~\ref{alg:state_preparation}.
\begin{algorithm}
    \caption{State preparation algorithm}
    \label{alg:state_preparation}

    \SetKwFunction{IsUniform}{IsUniform}
    \SetKwFunction{First}{First}
    \SetKwFunction{Subscript}{Subscript}
    \SetKwInOut{Input}{input}
    \SetKwInOut{Output}{output}
    \Input{A vector of real amplitudes $\psi$ of size $n \ge 1$ such that $\lVert\psi\rVert_2 = 1$}
    \Output{State preparation circuit}

    $d \gets n$\;
    $\Psi \gets \{\psi\}$\;

    \While{$d \ne 0$}{
        $\tilde\Psi \gets \varnothing$\;
        \tcc{Process each state $\psi$ in $\Psi$}
        \ForEach{$\psi \in \Psi$}{
            \If{$\text{IsUniform}(\psi)$ is False}{
                $j \gets \text{Subscript}(\text{First}(\psi))$\;
                \tcc{Split $\psi$ into two parts}
                $\tilde\Psi \gets \tilde\Psi \cup \{ (\psi_j, \dots, \psi_{j + 2^{d-1}-1}) \}$\;
                $\tilde\Psi \gets \tilde\Psi \cup \{ (\psi_{j + 2^{d-1}}, \dots, \psi_{j + 2^d - 1}) \}$\;
                \tcc{Compute rotation angle}
                $u \gets \sqrt{\sum_{k=j}^{j + 2^{d-1}-1} \psi_k^2}$\;
                $v \gets \sqrt{\sum_{k=j}^{j + 2^d - 1} \psi_k^2}$\;
                $\theta \gets 2\arccos(u/v)$\;
                \eIf{$d = n$}{
                    Apply MCRY gate with angle $\theta$ to qubit $(d - 1)$\;
                }{
                    Apply MCRY gate with angle $(\theta - \pi/2)$ to qubit $(d - 1)$ controlled by qubits $(n - 1)$ through $d$ in state $|\text{bin}(j)\rangle$\;
                }
            }
            \If{$d > 1$}{
                Apply Hadamard gate to qubit $(d - 2)$\;
            }
            $\Psi \gets \tilde\Psi$\;
            $d \gets d - 1$\;
        }
    }
\end{algorithm}

\subsubsection{Gate complexity estimate}
A circuit generated by Algorithm~\ref{alg:state_preparation} requires three types of components, namely, single-qubit $R_Y$, Hadamard $H$, and MCRY gates. From the algorithmic description of the procedure, it is clear that there is one single-qubit $R_Y$ rotation gate and $n - 1$ Hadamard gates, and their combined gate complexity is, therefore, ${\cal O}(n)$. The combined gate complexity of all MCRY gates is, however, less straightforward. On the one hand, there are $n - 1$ iterations of the outer loop in which MCRY gates might be applied. On the other hand, interpreting $|\psi\rangle$ as a piecewise-constant vector consisting of $p$ parts, the number of MCRY gates applied in each iteration is at most $p - 1$. This is due to the fact that starting from the second iteration of the outer loop of Algorithm~\ref{alg:state_preparation}, at most $p - 1$ non-uniform parts will be present in $\tilde \Psi$, each requiring one MCRY gate. Thus, there are at most $(p - 1)(n - 1)$ MCRY gates in total. A $k$-qubit MCRY gate can be implemented using ${\cal O}(k)$ CNOT and single-qubit gates \cite{zindorf2025efficient}. The gate complexity of any $k$-qubit MCRY gate in Algorithm~\ref{alg:state_preparation}, for simplicity, can be bounded from above by the complexity of the $n$-qubit MCRY gate, since $n \geq k$. Since there are now $(p - 1)(n - 1)$ MCRY gates, each requiring ${\cal O}(n)$ single-qubit and CNOT gates, the combined gate complexity of all MCRY gates is ${\cal O}(p n^2)$. Because the combined gate complexity of the other gates is only ${\cal O}(n)$, the gate complexity of the entire state preparation circuit is ${\cal O}(p n^2)$. This complexity is significantly lower than that of general state preparation methods, which typically have exponential complexity ${\cal O}(2^n)$.

It is worth pointing out, that the actual gate complexity also depends on how amplitudes in $|\psi\rangle$ are arranged and can be reduced drastically in special cases, e.g., when $|\psi\rangle$ is comprised of $p$ equally-sized parts, in which case the gate complexity can be reduced to ${\cal O}(p n)$. The gate complexity ${\cal O}(p n^2)$ obtained above corresponds to the worst case.

\section{Results}
In the following, we present the results of the VQA simulations. They are performed with the quantum simulation software Qiskit version 1 \cite{Qiskit} in ideal statevector and shot-based simulations applying 4 to 6 qubit ansatz circuits. Before we discuss the performance of the quantum algorithm, we investigate the expressibility of the ansatz circuit and the classical optimization algorithms which have been compared in our study. 

\subsection{Expressibility of parametric quantum circuit}
\begin{figure*}
    \centering
    \begin{minipage}[c]{0.48\textwidth}
        \includegraphics[width=\textwidth]{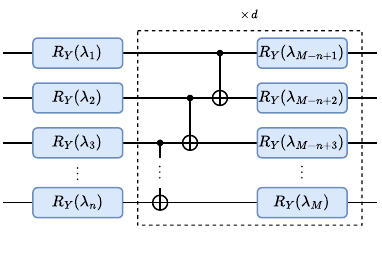}
    \end{minipage}
    \hfill
    \begin{minipage}[c]{0.48\textwidth}
        \includegraphics[width=\textwidth]{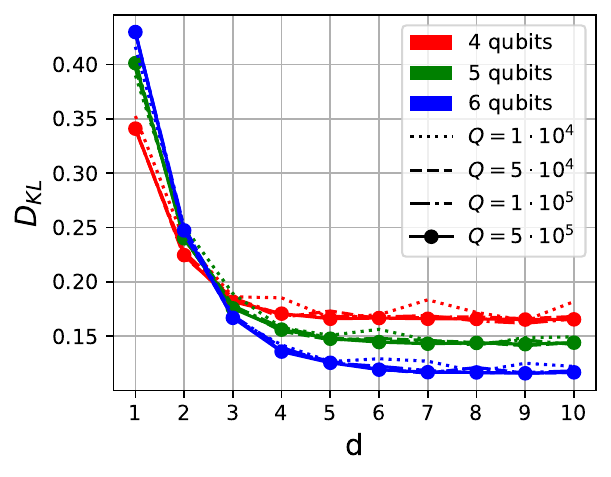}
    \end{minipage}
    \caption{Real-valued parametric quantum circuit (RPQC) and its expressibility in terms of the Kullback-Leibler divergence. Left panel shows the RPQC from the Qiskit circuit library with reverse linear entanglement and $d$ layers. Right panel shows expressibility of the RPQC as a function of the depth $d$ for 4-, 5-, and 6-qubit ansätze. The legend displays the number of pairs $Q$ that have been generated.}
    \label{fig:real_amp_ansatz_and_expressibility}
\end{figure*}

To encode the transient charge profile $\tilde c(x, t)$ in the VQA, the real-valued parametric quantum circuit (RPQC) with reverse linear entanglement was selected from the Qiskit circuit library. As depicted in Fig.~\ref{fig:real_amp_ansatz_and_expressibility}(a), this circuit consists of an initial layer of $R_Y$ gates followed by $d$ layers of alternating CNOT and $R_Y$ gates, resulting in a quantum state with real-valued probability amplitudes. Given $n$ qubits, the number of parameters in the RPQC is $n(d + 1)$.

We analyze the expressibility of the RPQC to find an optimal selection of its depth, $d$. The expressibility of a circuit is defined as the ability to explore the Hilbert space through an entangled unitary circuit. In general, it can be quantified using the Kullback-Leibler divergence $D_{\text{KL}}$ of two probability density functions (PDF) \cite{sim2019expressibility}, the distribution of fidelities $0 \leq F \leq 1$ of Haar random states $P_{\text{Haar}}(F)$ and the distribution of $F$ resulting from the PQC, denoted as $P_{\text{RPQC}}(F; \bm \lambda)$. For the ensemble of Haar random states, the analytical form of the PDF of fidelities \cite{zyczkowski2005average} is $P_{\text{Haar}}(F) = (N - 1)(1 - F)^{N - 2}$ with $N = 2^n$. Therefore, the measure of expressibility is:
\begin{equation}
    D_{\text{KL}} = \int_0^1 P_{\text{RPQC}}(F; \bm \lambda) \log\left(\frac{P_{\text{RPQC}}(F; \bm \lambda)}{P_{\text{Haar}}(F)}\right)dF.
\end{equation}
Since Haar random states uniformly sample the Hilbert space of pure states, the smaller the $D_{\text{KL}}$, the better the states generated by the RPQC approximate this uniform distribution over the space, and therefore, the higher the expressibility of the RPQC. The $P_{\text{RPQC}}(F; \bm \lambda)$ PDF is calculated as follows (see also ref. \cite{Koecher2025}): (1) $Q$ pairs of parameter vectors $\bm \lambda_1$ and $\bm \lambda_2$ are sampled uniformly and independently; (2) the RPQC is used to generate the corresponding states $|\psi(\bm \lambda_1)\rangle = U(\bm \lambda_1)|0\rangle$ and $|\psi(\bm \lambda_2)\rangle = U(\bm \lambda_2)|0\rangle$; (3) their fidelities $F = \lvert\langle\psi(\bm \lambda_1) \vert \psi(\bm \lambda_2)\rangle\rvert^2$ are calculated; (4) $P_{\text{RPQC}}(F; \bm \lambda)$ is approximated using the probability density histogram of the previously obtained fidelities $F$. Figure~\ref{fig:real_amp_ansatz_and_expressibility}(b) displays the results for $D_{\text{KL}}$ for the 6-qubit RPQC as a function of the circuit depth $d$. The number of bins used to obtain the probability density histogram is 320. The number of quantum state pairs is given in the legend. It can be seen that $D_{\text{KL}}$ decays with increasing $d$ and saturates for $d \geq 4$, $d \geq 5$ and $d \geq 6$ for 4-, 5- and 6-qubit ansätze respectively. The number of pairs, $Q$, has been varied from $10^4$ to $5 \times 10^5$, which did not change the result. We use $D_\text{KL}$ as a general measure, for simplicity. The ansatz circuit is thus a problem-agnostic choice and does not incorporate application-specific properties.

\subsection{Circuit depth estimate of VQA}
Evaluating the VQA cost function relies on executing four circuits corresponding to the four terms $S_\text{LIN}$, $S^\pm$, $S_\text{PER}$ and $S_\text{BND}$. These circuits follow a Hadamard test architecture, which is generally depth-intensive. In the following, we provide circuit depth estimates for these cost-evaluation circuits.

The ansatz circuit used in this work consists of repetitions of entangling layers and $R_Y$ gates, and its circuit depth scales as ${\cal O}(dn)$. The circuit for evaluating the term $S_\text{LIN}$ is a Hadamard test circuit that employs controlled versions of the ansatz, where $R_Y$ are replaced by controlled $R_Y$ (CRY) gates and CNOT gates by Toffoli gates (also known as CCNOT gates). Considering the standard decompositions of these gate types, the estimated depth of $S_\text{LIN}$ is ${\cal O}(dn + n)$.

The $S^\pm$ circuit incorporates the ansatz circuit in an uncontrolled form, a controlled state preparation circuit for encoding the coefficients, and $n$ Toffoli gates. When the state preparation circuit becomes controlled, its MCRY gates gain an additional control qubit, increasing the depth from ${\cal O}(pn^2)$ to ${\cal O}(pn^2 + pn)$. Consequently, the total circuit depth scales as ${\cal O}[pn^2 + (p + d + 1)n]$. See again subsection III.B.2 with $p$ being the number of the piecewise constant state vector parts.

The $S_\text{PER}$ circuit is similar to $S^\pm$, except that it additionally includes the shift block $\hat S_-$, which consists of $n$ multi-qubit Toffoli gates. Since an $n$-qubit Toffoli can be decomposed with depth scaling linearly \cite{daSilva2022}, the depth of the shift block $\hat S_-$ scales as ${\cal O}(n^2)$. Consequently, the overall circuit depth of $S_\text{PER}$ scales as ${\cal O}[(p + 1)n^2 + (p + d + 1)n]$. The circuit used to evaluate the corrective term $S_\text{BND}$, proposed by Over et al. \cite{over2025boundary}, also relies on the shift block $\hat S_-$, and, therefore, exhibits a depth scaling of ${\cal O}(n^2)$.

Combining all the estimates above, the total depth estimate per cost function evaluation is ${\cal O}[(p + 1)n^2 + (p + d + 1)n]$. This scaling is dominated by the evaluation of the $S_\text{PER}$ term, which represents the most computationally expensive component. Although the shift block $\hat S_-$ can be implemented in principle with sublinear depth scaling, e.g. by introducing ancillary qubits, the overall scaling of the algorithm remains quadratic in $n$ due to the state preparation required for encoding the coefficients.

\subsection{Adaption of surrogate-based optimization}
The classical optimization algorithms, such as the Nelder–Mead (NM) \cite{Nelder1965}, the Broyden-Fletcher-Goldfarb-Shanno (BFGS), covariance matrix adaptation evolution strategy (CMA-ES) \cite{Hansen2023}, and a surrogate-based optimization with kernel approximations (SBO) \cite{Shaffer2023}, were used for detecting the minimum of the VQA cost function.

\begin{figure}
     \centering
     \includegraphics[width=.4\textwidth]{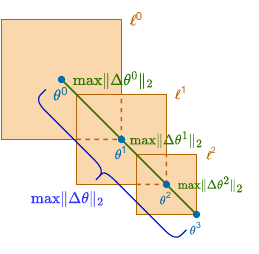}
     \caption{Sketch of the maximum distance $\max \|\Delta \bm \theta \|_2$ that can be covered by the surrogate-based optimization (SBO).}
     \label{fig:sbo_patch_size}
\end{figure}

The idea behind the latter SBO algorithm is to approximate the cost function based on its $\tau$ random samples within a hypercube of length $\ell^0$ centered around an initial guess ${\bm \theta}^0 = (\theta_0^0, \theta_1^0, \dots, \theta_M^0)$ of the variational parameters. The obtained approximation is then optimized classically subject to the box-constraints corresponding to the hypercube. Once the local minimum is found, it becomes the center of the next hypercube of smaller length $\ell^1$ and the process repeats. The algorithm requires a user to provide parameter values including among others the size of the initial hypercube $\ell^0$, the number of points $\tau$ to sample within each hypercube, and the number of iterations $n_\text{iter}$. As suggested by the original paper, $\tau$ is kept fixed. For setting $n_\text{iter}$, we use a simple heuristic formula,
\begin{equation}
    n_\text{iter} = \left\lceil \frac{n_\text{fev}}{\tau} \right\rceil\,,
\end{equation}
where $n_\text{fev}$ is the overall function evaluation budget. For setting the initial hypercube length $\ell^0$, we use two approaches. In the first, $\ell^0$ is set to $\| \bm \theta_1 - \bm \theta_0 \|_\infty$ for all time steps, meaning that the hypercube contains the solution to the next time step, since $\| \bm \theta_1 - \bm \theta_0 \|_\infty \geq \| \bm \theta_{j + 1} - \bm \theta_j \|_\infty$ assuming that RPQC is expressible enough. We call this approach fixed patch size (FPS). The second approach, called heuristic patch size (HPS), sets the patch size according to the number of iterations $n_{\text{iter}}$ and the number of variational parameters $M + 1$ such that the maximum Euclidean distance that the algorithm can cover within $n_{\text{iter}}$ iterations is proportional to $\lVert \bm \theta_1 - \bm \theta_0 \rVert_2$. That is, in each iteration of the optimization algorithm the maximum distance covered by the algorithm is a half diagonal of the hypercube, as depicted in Fig.~\ref{fig:sbo_patch_size}, and thus
\begin{equation}
    \max \| \Delta \bm \theta^i \| = \frac{\sqrt{M + 1}}{2}\ell^i\,.
\end{equation}
Taking into account that $\ell^i = \ell^0(1 - i / n_\text{iter})$, the total maximum distance covered by the algorithm is given by
\begin{equation}
    \max \|\Delta \bm \theta \|_2 = \frac{\sqrt{M + 1} (n_\text{iter} + 1)}{4}\,\ell^0\,.
\end{equation}
Therefore, requiring that $\max \|\Delta\bm\theta\|_2 = \|\bm\theta_1 - \bm\theta_0\|_2$, the initial patch size can now be defined in terms of the maximum distance between two consecutive variational parameter sets:
\begin{equation}
    \ell^0 = \frac{4(\gamma + 1)\|\bm \theta_1 - \bm \theta_0\|_2}{\sqrt{M + 1} (n_\text{iter} + 1)}\,,
\end{equation}
where $\gamma \geq 0$ is an empirical parameter that accounts for, e.g., zigzag behavior of the iterates.

The vector of the variational parameters $\bm \theta$ used in the SBO algorithm contains $M + 1$ angles in the range $[0, 2\pi]$, where the first angle should correspond to the normalization constant $\lambda_0$ from the VQA formulation. To map the variational parameters $(\lambda_0, \bm \lambda)$ used in \eqref{eq:lambda_0} and \eqref{eq:c_hat_l} to $\bm \theta$, we use the following relations:
\begin{align}
    \theta_0 = 2\pi\frac{\lambda_0^l}{\lambda_0^0} \quad\mbox{and}\quad
    \theta_j = \lambda_j^l,\quad j = 1, \dots, M\,,
\end{align}
where $\lambda_0^0$ and $\lambda_0^l$ are normalization constants of the initial conditions and the solution at $l$-th time step, respectively. Since the VQA simulates the transient solution $\tilde c(x, t)$, its 2-norm, or, equivalently, the normalization constant $\lambda_0^l$, is bounded from above by $\lambda_0^0$ and tends to converge to zero as $t \rightarrow \infty$, therefore, $\theta_0 \in [0, 2\pi]$, as required by the SBO algorithm.

\begin{figure*}[t!]
     \centering
     \includegraphics[width=.9\textwidth]{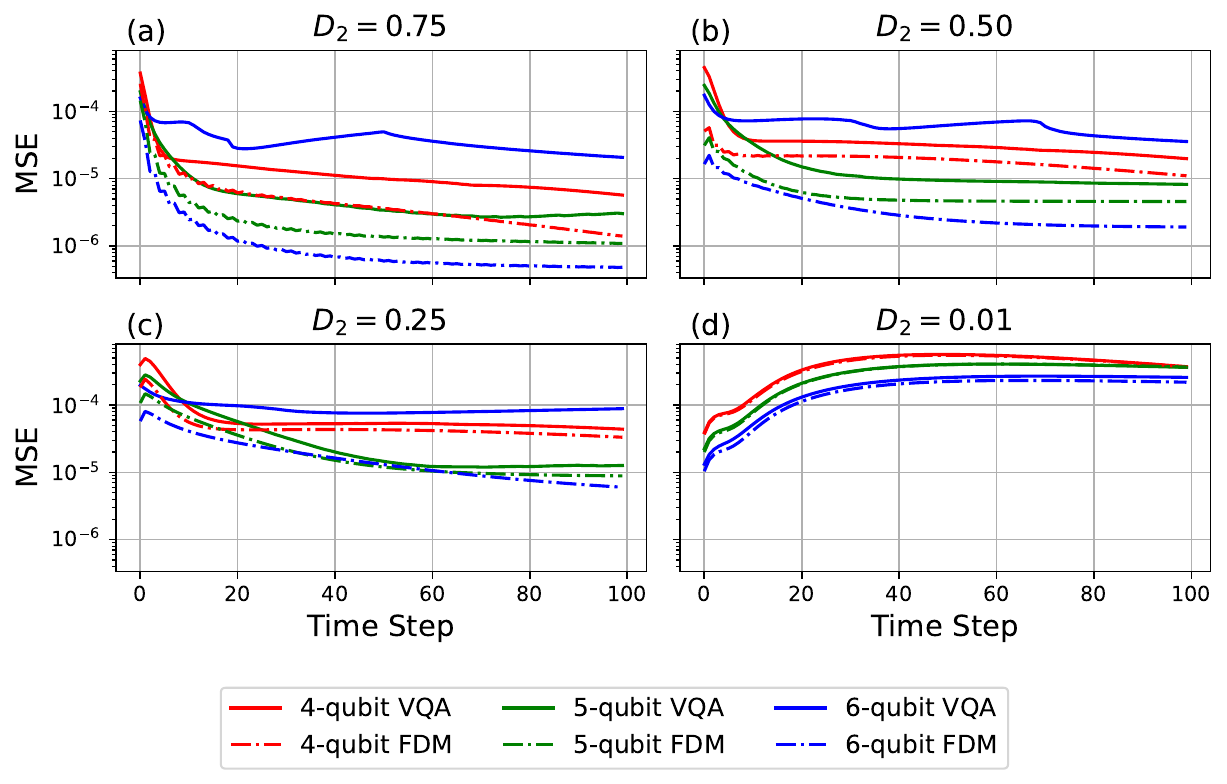}
     \caption{Mean squared error (MSE) versus time step of the FDM and VQA solutions relative to the analytical solution with $D_1 = 1$ and $D_2 = 0.01, 0.25, 0.50, 0.75$ obtained with 4-, 5- and 6-qubit statevector simulations.}
     \label{fig:as_vs_fdm_vqa_mse}
\end{figure*}
\begin{figure}[t!]
     \centering
     \includegraphics[width=.48\textwidth]{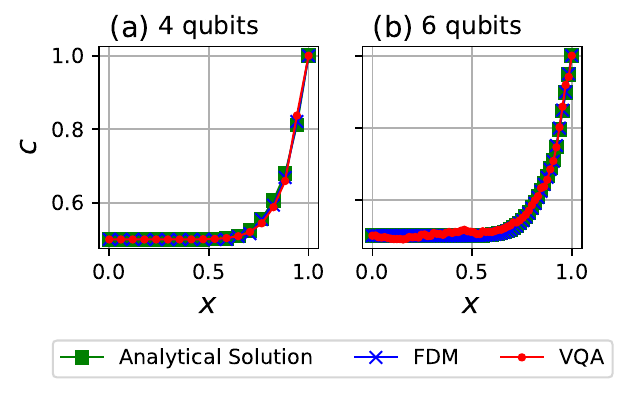}
     \caption{Concentration profile for $D_2 = 0.5$ at $t = 8 \times 10^{-3}$ for (a) 4 qubits and (b) 6 qubits.}
     \label{fig:4_vs_6_qubits}
\end{figure}
\begin{figure*}
     \centering
     \includegraphics[width=.95\textwidth]{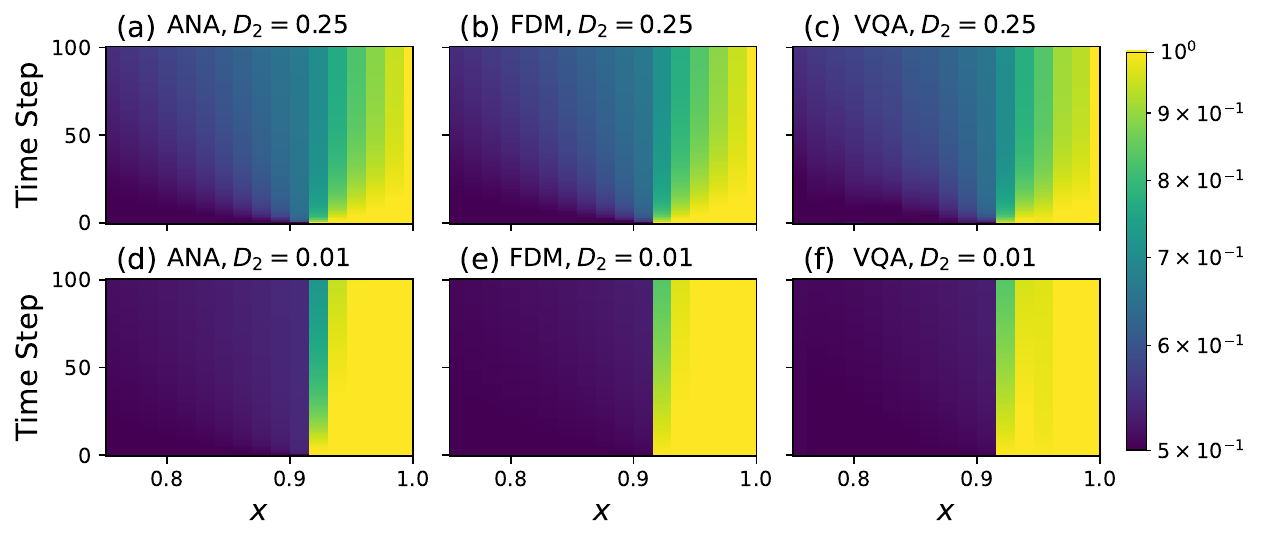}
     \caption{Comparison between the time evolutions shown in panels (a)--(c) for $D_2 = 0.25$ and panels (d)--(f) for $D_2 = 0.01$. The number of internal grid points used in the simulations is $N = 64$. The $x$-axis is restricted to the range $x \in [0.75, 1]$, since the concentration is nearly zero for $x < 0.75$.}
     \label{fig:0.25_vs_0.01}
\end{figure*}

\subsection{Time evolution of concentration profile}
\subsubsection{Quantum simulation parameters}
The diffusion equation \eqref{eq:diffusion_equation} was solved approximately for 100 time steps with an integration step width $\Delta t = \Delta x^2 / 2$ using the FDM and VQA methods with $D_1 = 1$, $D_2 = 0.01, 0.25, 0.50, 0.75$ and $N = 16, 32, 64$ with ideal state-vector simulations. The interface position $\tilde x_1=10 / 11$ as in Fig.~\ref{fig:aem_problem}(b), with $\tilde{x}_0 = 0$ and $\tilde{x}_2 = 1$ (recall that all physical quantities are already non-dimensionalized). Ansatz circuits for the real amplitudes with $d = 4$ layers in case of 4-qubit, $d = 5$ layers in case of 5-qubit, and $d = 6$ layers in case of 6-qubit experiments were used. The transient solution $\tilde c(x, t)$ is combined with the steady state solution $c_s(x)$ as prescribed by \eqref{eq:ss_ts_decomposition} to produce the full ion concentration profile $c(x, t)$. This is then compared with the FDM result and the analytical solution. Recall also that the analytical series solution is obtained using 30000 terms in eq. \eqref{eq:analytical_solution}. 

The BFGS algorithm was chosen at the beginning as the optimization routine, with a maximum number of iterations of 100 and $\lVert\nabla C(\theta)\rVert_2 < 1 \times 10^{-3}$ as the termination criterion. Figure~\ref{fig:as_vs_fdm_vqa_mse} shows the mean squared error (MSE) of the FDM and VQA methods relative to the analytical solution \eqref{eq:analytical_solution}.

Coefficient vectors of the cost function terms $S_\text{PER}$ and $S^\pm$ for $D_2 = 0.50$, along with gate counts and circuit depths for preparing these coefficients, are reported in Table I. Circuits for preparing coefficients of $S^\pm$ are deeper because their coefficient vectors consist of 3 constant parts, unlike those of $S_\text{PER}$, which have only 2 parts. Note that the structure of the coefficient vectors, the gate count and the circuit depth of the corresponding state preparation circuits are independent of $D_2$. They depend on the interface position $\tilde{x}_1$ only. That is, changing $D_2$ affects the prefactors of the constant parts only, not their arrangement in the coefficient vector.

\begin{table*}
    \label{tab:coefficients_depths}
    \centering
    \begin{tabular}{ccccccc}
        \hline\hline
        No. of qubits $\;$ & Coefficients in $S_\text{PER}$ $\;$ & No. of gates $\;$ & Depth $\;$ & Coefficients in $S^\pm$ $\;$ & No. of gates $\;$ & Depth $\;$ \\
        \hline
        4 & $\underbrace{0.26, \dots, 0.26}_{\#14}, 0.13, 0.13$ & 10 & 7 & $\underbrace{0.26, \dots, 0.26}_{\#14}, 0.19, 0.13$ & 23 & 20 \\
        5 & $\underbrace{0.19, \dots, 0.19}_{\#28}, \underbrace{0.09, \dots, 0.09}_{\#4}$ & 11 & 7 & $\underbrace{0.18, \dots, 0.18}_{\#28}, 0.14, 0.09, 0.09, 0.09$ & 34 & 28 \\
        6 & $\underbrace{0.13, \dots, 0.13}_{\#58}, \underbrace{0.06, \dots, 0.06}_{\#6}$ & 35 & 28 & $\underbrace{0.13, \dots, 0.13}_{\#58}, 0.10, \underbrace{0.06, \dots, 0.06}_{\#5}$ & 105 & 81 \\
        \hline\hline
    \end{tabular}
    \caption{Coefficients in the cost function terms $S_\text{PER}$ and $S^\pm$ for $D_2 = 0.50$, together with gate counts and circuit depths of the corresponding state preparation circuits for $n = 4, 5$ and $6$ qubits. The structure of the coefficient vectors, gate counts and circuit depths are independent of the value of $D_2$.}
\end{table*}

\subsubsection{Time evolution and dependence on strength of discontinuity of diffusion constant}
Inspecting the graphs, one can deduce that, in terms of MSE, the FDM always performs better than the VQA. The reason is that while the VQA utilizes the same finite differences as the FDM, it uses an ansatz circuit to express the solution, which inevitably introduces error and steers the solution away from the FDM. Moreover, increasing the spatial resolution leads to a higher MSE of the VQA as opposed to the FDM when 6 qubits are employed as can be seen in Figs.~\ref{fig:as_vs_fdm_vqa_mse}(a)--(c). This is because even though 6 layers achieve the maximum expressibility of a 6-qubit RPQC, the number of free parameters in this case is still less than the number of grid points, which leads to the concentrations at grid points being under-determined. Figure~\ref{fig:4_vs_6_qubits} showing the concentration profile at $t = 8 \times 10^{-3}$ for $D_2 = 0.5$ illustrates this point. Nevertheless, the capability of the VQA method to numerically solve this specific problem is demonstrated. When the number of qubits is $n = 6$, the charge profile shown in Fig.~\ref{fig:4_vs_6_qubits}(b) exhibits small spatial oscillations in the region $x \in [0, 0.75]$, which are absent in the case of $n = 4$ depicted in Fig.~\ref{fig:4_vs_6_qubits}(a).

The MSE curves for $D_2 = 0.01$, which are depicted in Fig.~\ref{fig:as_vs_fdm_vqa_mse}(d), differ significantly from other cases due to a monotonic increase in MSE in time steps $l \in [0, 40]$. This difference can be explained in parts by comparing the charge profiles for $D_2 = 0.25$ and $D_2 = 0.01$ resolved using $N = 64$ internal grid points, as shown in Fig.~\ref{fig:0.25_vs_0.01}. In both cases, the initial charge distributions are piecewise constant profiles, with a sharp change at the interface $\tilde x_1$ between the two layers. For $D_2 = 0.25$, this sharp transition is maintained for approximately 5 time steps before smoothing out. This behavior corresponds to the short-term increase followed by a decrease in MSE near time step $l = 0$ observed in Fig.~\ref{fig:as_vs_fdm_vqa_mse}(c). 

A similar trend is also evident for MSE in the case of $D_2 = 0.50$ shown in Fig.~\ref{fig:as_vs_fdm_vqa_mse}(b). However, when $D_2 = 0.01$, the sharp transition in the charge profile is maintained throughout the entire duration of the simulation. Similarly to the cases $D_2 = 0.50$ and $D_2 = 0.25$, the corresponding MSE curves in Fig.~\ref{fig:as_vs_fdm_vqa_mse}(d) exhibit an increase, but this increase occurs over a much longer timescale. A decrease in MSE is observed only for $n = 4$, where the larger time step allows the simulation to progress into the phase where the initially sharp transition begins to smooth out. Therefore, the lower the value of $D_2$, the more pronounced and prolonged the initial increase in the MSE becomes. This observation is consistent with the fact that the relaxation rate of the system decreases as $D_2$ decreases, as illustrated in Fig.~\ref{fig:relaxation_rate_analysis}(b). A slower relaxation rate means that the sharp transition in the charge profile diffuses more slowly, prolonging the duration of elevated MSE, as is the case for $D_2 = 0.01$.

Apart from that, when $D_2 = 0.01$, the performance of the VQA closely resembles that of the FDM, with increasing spatial resolution leading to improved overall accuracy, even when employing 6 qubits. One can notice that the MSE curves of the FDM and VQA for the same number of qubits converge as $D_2$ decreases, as shown in Fig.~\ref{fig:as_vs_fdm_vqa_mse}. Moreover, the MSE curves of the FDM tend to increase as $D_2$ decreases. Since the VQA uses the same spatial discretization as the FDM, this suggests that for $D_2 \rightarrow 0$, the MSE of the VQA is dominated by discretization errors rather than limitations of the PQC in representing the charge profile. It also explains why the VQA performs comparatively well in the low-diffusivity regime of the membrane.

\begin{figure}
     \centering
     \includegraphics[width=.48\textwidth]{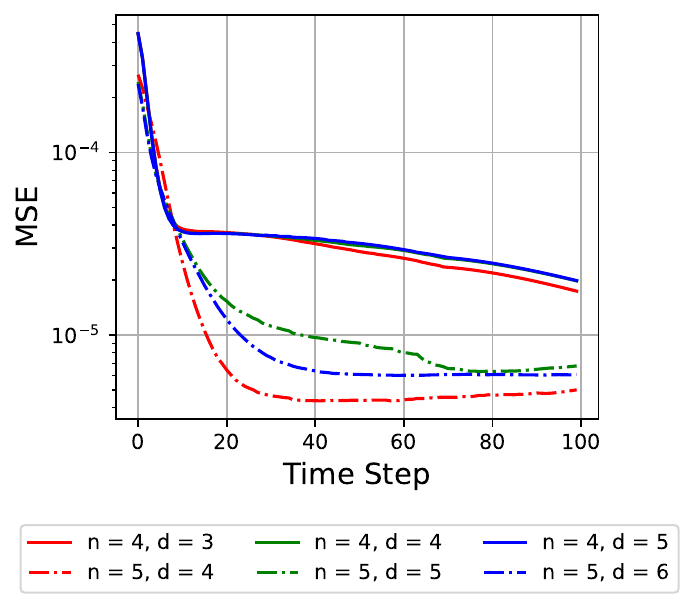}
     \caption{Mean squared error (MSE) verus time step of the VQA solutions relative to the analytical solution for $D_2 = 0.50$, obtained with $n = 4$ and $5$ qubits and various numbers of layers $d$ (see legend) in the RPQC.}
     \label{fig:vqa_vs_as_mse_45_345_456_0.50}
\end{figure}

The number of layers $d$ in the RPQC was chosen to ensure high expressibility. Specifically, values of $d$ at which the Kullback-Leibler divergence begins to saturate were selected, resulting in, for example, $d = 4$ for $n = 4$ qubits and $d = 5$ for $n = 5$ qubits. We verified that slightly increasing or decreasing $d$ at fixed qubit number does not significantly affect the accuracy of the algorithm, but increases its runtime. Figure~\ref{fig:vqa_vs_as_mse_45_345_456_0.50} shows MSE curves for $n = 4, 5$ and various values of $d$ when $D_2 = 0.50$. As seen, the largest effect on accuracy stems from the number of qubits $n$ rather than the number of layers $d$. In fact, increasing $d$ does not necessarily improve accuracy. This behavior arises because accuracy depends not only on $n$ and $d$, but also on optimization parameters such as the gradient-norm tolerance in the BFGS algorithm, maximum number of iterations, etc. These parameters must be adjusted since changing $d$ alters the optimization landscape. Table II reports the runtime of the algorithm with respect to $n$ and $d$, showing that increasing $d$ leads to longer run times.

\begin{table}
    \label{tab:computation_time_vqa_45_345_456}
    \centering
    \begin{tabular}{ccc}
        \hline\hline
        No. of qubits $n\;\;$ & No. of layers $d\;\;$ & Time \\
        \hline
        \multirow{4}{*}{4} & 3 &  0.15 h \\
         & 4 &  0.21 h \\
         & 5 &  0.25 h \\ \bottomrule
        \multirow{4}{*}{5} & 4 &  1.28 h \\
         & 5 &  2.00 h \\
         & 6 &  2.09 h \\ \bottomrule
        \hline\hline
    \end{tabular}
    \caption{Comparison of computation time of the algorithm for different qubit numbers $n$ and RPQC layer numbers $d$ when $D_2 = 0.50$. All ideal statevector simulations are done with the BFGS algorithm.}
\end{table}

\subsubsection{Results for different optimization schemes}
\begin{figure}
     \centering
     \includegraphics[width=.48\textwidth]{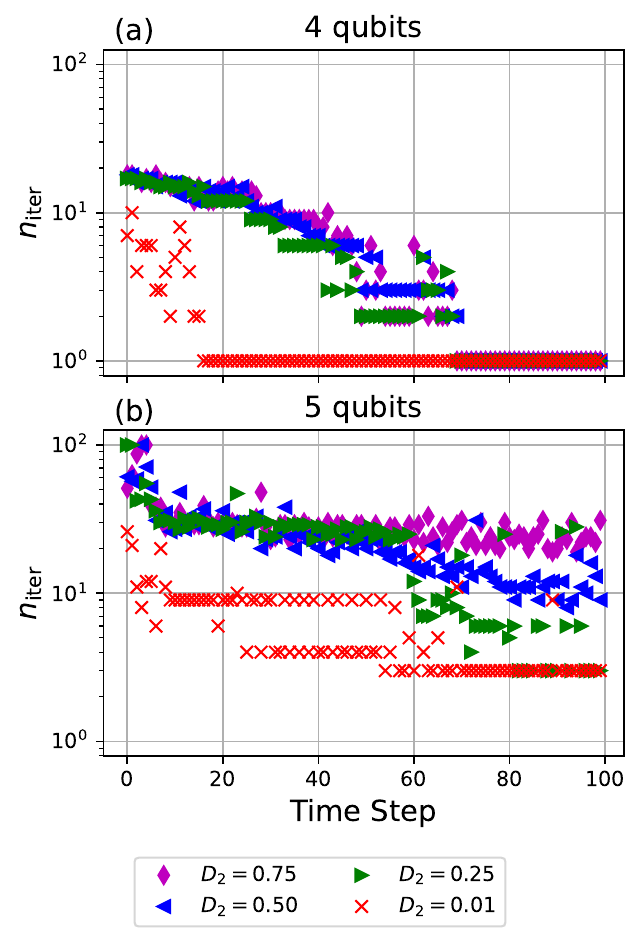}
     \caption{Number of iterations of the BFGS optimization algorithm in the VQA for (a) 4 qubits and (b) 5 qubits.}
     \label{fig:n_iter_bfgs}
\end{figure}

The number of iterations of the BFGS algorithm used in the VQA for various numbers of qubits and different values of $D_2$ is shown in Figs.~\ref{fig:n_iter_bfgs}(a,b). As shown, when $D_2 = 0.01$, the BFGS algorithm requires the fewest iterations compared to other cases. For example, Fig.~\ref{fig:n_iter_bfgs}(a) illustrates that after approximately 15 time steps ($l=15$), a single iteration of the BFGS algorithm suffices to achieve the required gradient tolerance of the cost function. Moreover, Figs.~\ref{fig:n_iter_bfgs}(a) and (b) indicate that more iterations are necessary at earlier times, near the initial conditions, compared to later times.
Since the relaxation rate decreases as $D_2 \rightarrow 0$, the spatial charge concentration profile $c(x, t)$ changes very slowly in time. Apart from that, the closer the discrete charge concentration profiles $c^l$ at consecutive time steps, the closer the corresponding variational parameters $\bm \lambda^l$ of the VQA ansatz. Therefore, when $D_2 = 0.01$, the variational parameters change moderately, resulting in a relatively low number of iterations required by the optimization routine. Table III shows the computation time used in the BFGS algorithm in the aforementioned experiments. As is evident, the computation time increases with both the number of qubits and the diffusivity of the membrane.
\begin{table}
    \label{tab:computation_time}
    \centering
    \begin{tabular}{ccc}
        \hline\hline
        No. of qubits $n\;\;$ & AEM diffusivity $D_2\;\;$ & Time \\
        \hline
        \multirow{4}{*}{4} & 0.01 &  0.18 h \\
         & 0.25 &  0.40 h \\
         & 0.50 &  0.41 h \\
         & 0.75 &  0.46 h \\ \bottomrule
        \multirow{4}{*}{5} & 0.01 &  1.55 h \\
         & 0.25 &  3.88 h \\
         & 0.50 &  3.62 h \\
         & 0.75 &  4.52 h \\ \bottomrule
        \multirow{4}{*}{6} & 0.01 & 25.44 h \\
         & 0.25 & 76.51 h \\
         & 0.50 & 76.02 h \\
         & 0.75 & 76.21 h \\
        \hline\hline
    \end{tabular}
    \caption{Comparison of computation time of the algorithm for different qubit numbers $n$ and diffusivity constants $D_2$. All ideal statevector simulations are done with the BFGS algorithm.} 
\end{table}

Large computation times presented in Table III have a dependence on the number of qubits $n$, but this dependence is not straightforward. Increasing $n$ leads to a dimension of the solution space growing as $N = 2^n$, which requires the ansatz circuit to encode solutions in higher-dimensional spaces. The simplest approach allowing actual solutions of the problem to be discovered by the ansatz circuit is to increase its depth thus improving its expressibility. However, as the ansatz circuit depth grows, the gradients of the VQA cost function vanish exponentially, requiring more computational resources to estimate the cost function accurately \cite{liu2023training}. Consequently, the optimization process is dominated by statevector simulations as shown in Table IV, as the increasingly flat cost function forces the optimization algorithm to perform numerous evaluations to identify reliable search directions. Therefore, the actual bottleneck is not the classical optimization algorithm, but the ansatz circuit itself. Moreover, restricting oneself only to gradient-free approaches does not contribute to mitigating the problem, as the issue affects both gradient-free and gradient-based optimization algorithms \cite{arrasmith2021effect}. However, there exist approaches that take the structure of the parameterized quantum circuit into account, thus improving convergence and mitigating barren plateaus in some cases \cite{Stokes2020, Gacon2021, Fitzek2024}.
\begin{table}
    \label{tab:computation_time_2}
    \centering
    \begin{tabular}{cccc}
        \hline\hline
        No. of qubits $n\;\;$ & BFGS $\;\;$ & Statevector $\;\;$ & Total $\;\;$\\
        \hline
        4 & 2 s & 280 s & 282 s \\
        5 & 7 s & 1953 s & 1960 s \\
        6 & 29 s & 15913 s & 15942 s \\
        \hline\hline
    \end{tabular}
    \caption{Comparison of computation times of different parts of the VQA algorithm for different qubit numbers $n$. The computation times for the statevector simulations and the calculations performed solely by the BFGS algorithm were measured separately. In all 4-, 5- and 6-qubit experiments the equation was solved for 10 time steps.}
\end{table}

\begin{figure}
     \centering
     \includegraphics[width=.48\textwidth]{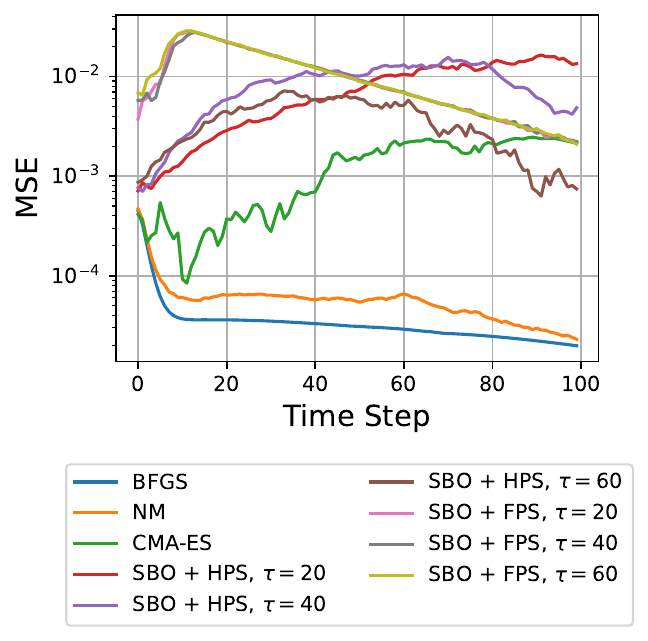}
     \caption{Comparison of various classical optimization algorithms for the case with $D_2 = 0.5$ using $N = 16$ (4 qubits) internal grid points. The legend indicates the specific method.}
     \label{fig:4_0.50_4_bfgs_nm_sbo_mse}
\end{figure}

We also compared various optimization routines for $n = 4$ and $D_2 = 0.5$, namely, the BFGS algorithm, the Nelder–Mead (NM) algorithm, the CMA-ES, and the SBO algorithm. For the SBO algorithm, we used both the FPS and HPS approaches and varied the number of sample points $\tau$. In the HPS method, the empirical parameter $\gamma$ was set to 1, effectively doubling the anticipated distance between variational parameters. Other parameters, such as $\epsilon_i$, $\epsilon_\text{int}$ and $\epsilon_f$ where set to $0$, $\ell^0 / 20$ and $\ell^0 / 2$ respectively. To ensure a fair comparison, all optimization algorithms employed the same number of function evaluations, $n_\text{fev}$. The results are presented in Fig.~\ref{fig:4_0.50_4_bfgs_nm_sbo_mse}.
The BFGS and NM algorithms achieved similar performance in terms of MSE, both yielding values on the order of $10^{-4}$, with NM producing marginally higher errors. CMA-ES and SBO algorithms produced MSE values of the order of $10^{-2}$ and $10^{-1}$ to $10^{-2}$, respectively, worse than those achieved by BFGS and NM. When the FPS approach is employed for the SBO algorithm, the corresponding MSE does not improve when increasing the number of sample points. However, when using the HPS approach, increasing the number of sample points reduces the error to ${\cal O}(10^{-2})$.

We also conducted 5-qubit statevector simulations using the aforementioned algorithms, wherein only the HPS approach was utilized for the SBO algorithm. In these tests, the BFGS algorithm demonstrated better performance in comparison to 4 qubits and convergent behavior in terms of MSE, whereas both the NM and SBO algorithms exhibited a stronger increasing MSE. As all algorithms were allocated with the same number of function evaluations, these results indicate that NM and SBO require more function evaluations to reach higher accuracy compared to BFGS. The analysis demonstrates again one weakness of the VQA methods, the choice and tuning of the classical optimization procedure for the non-convex minimum search of the cost function is task-specific.

\begin{figure*}
     \centering
     \includegraphics[width=.95\textwidth]{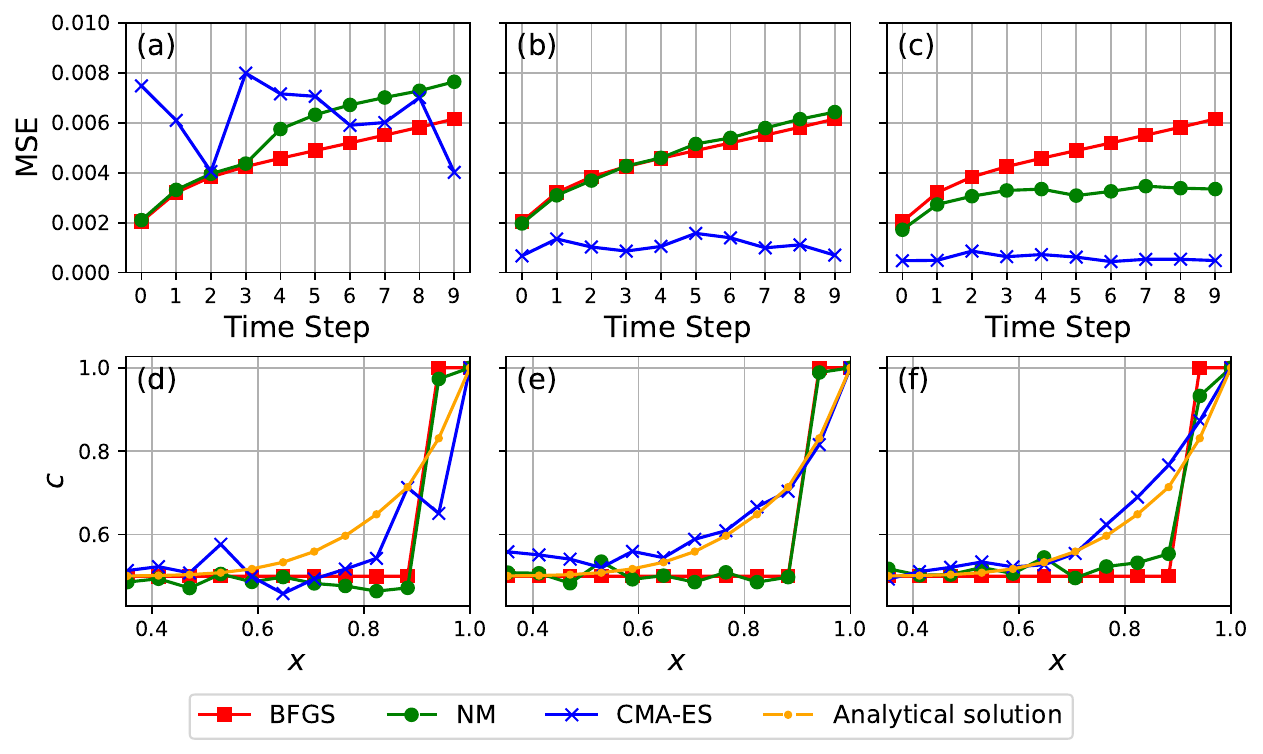}
     \caption{Comparison of various classical optimization algorithms subject to measurement noise for the case with $D_2 = 0.5$ using $N = 16$ (4 qubits) internal grid points. Panels (a)--(c) show mean squared error (MSE) versus time step of the VQA relative to the analytical solution when $10^4$, $10^5$ and $10^6$ shots are employed, respectively. Panels (d)--(f) show respective concentration profiles at $t = 1.7 \times 10^{-4}$ corresponding to 10 time steps. The legend holds for all panels.}
     \label{fig:4_0.50_4_sb_bfgs_nm_cma_mse}
\end{figure*}

Apart from ideal statevector simulations, shot-based simulations were also performed for the aforementioned 4-qubit example to examine the algorithm's robustness to sampling noise arising from the measurement process. There, only three optimization routines (BFGS, NM and CMA-ES) were employed to simulate the evolution over 10 time steps. To examine the effect of the number of shots, three values were considered, $10^4$, $10^5$ and $10^6$. The corresponding results are presented in Fig.~\ref{fig:4_0.50_4_sb_bfgs_nm_cma_mse}.
The BFGS and NM algorithms failed to advance the evolution of the initial conditions. As shown, in both cases the concentration remained close to their initial conditions. Moreover, increasing the number of shots did not significantly improve the performance of either algorithm. In contrast, starting from $10^5$ shots, CMA-ES achieved good agreement with the analytical solution, where higher shot counts leading to a lower overall MSE. The results demonstrate that conventional optimization routines, such as BFGS and NM, are suitable only for ideal statevector simulation scenarios, as they are highly sensitive to sampling noise. The success of CMA-ES can be attributed to its inherent robustness to noise in the cost function, making it a promising candidate for use in VQAs on real hardware \cite{Novak2025}. This study confirms that the best performing optimization scheme is problem-specific, cf. ref.\cite{Ingelmann2024}.

\begin{figure}
     \centering
     \includegraphics[width=.45\textwidth]{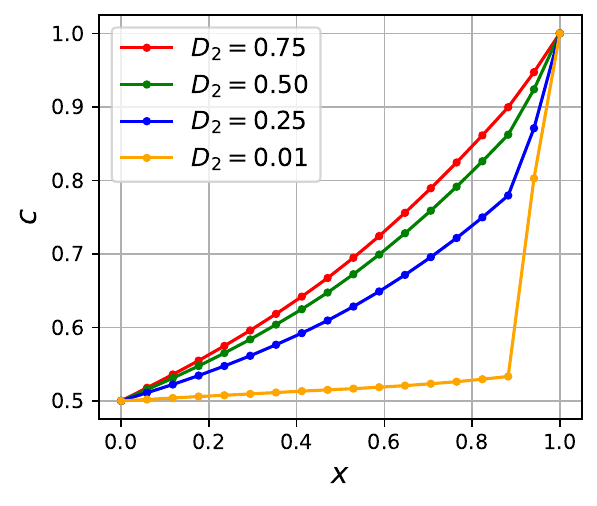}
     \caption{Comparison of the anion concentration profiles after 100 integration time steps for various values of $D_2$. The number of internal grid points used in the VQA simulations is $N = 16$. All runs are relaxed already close to the steady state.}
     \label{fig:steady_state_ana_vs_vqa_2}
\end{figure}


\section{Conclusion}
The objective of the present work is to investigate the diffusive ion transport problem across a multi-layered anion exchange membrane (AEM) and simulate it using a quantum algorithm. The anion transport problem is formulated as a time-dependent linear and one-dimensional diffusion problem with a space-dependent diffusion coefficient and inhomogeneous Dirichlet boundary conditions from a macroscopic perspective. The central goal is to study the concentration contrasts, which can build up during operation of electrolyzers. This transport process can be modeled as a one-dimensional problem which is suited for a quantum algorithm with a relatively small number of qubits. This is a second motivation for our work.

The quantum algorithm used for solving the diffusion problem is a variational quantum algorithm (VQA) based on the weak formulation framework. Since the problem does not represent a unitary time evolution and is subject to non-periodic boundary conditions, it requires a specific quantum algorithm implementation with corresponding boundary treatment, similar to \cite{over2025boundary}. We have investigated its performance under ideal statevector simulations using $N = 16, 32, 64$ spatial grid points corresponding to 4, 5, and 6 data qubits, respectively. We also demonstrated that shot-based simulations are applicable. The deviation from the analytical solution decreases with increasing number of shots (and thus time complexity). Moreover, we simultaneously varied the diffusivity of the AEM from $D_2 = 0.01$ to $0.75$ while keeping $D_1 = 1$ fixed. This varies the magnitude of the discontinuity inside the two-layer membrane. The algorithm's performance was measured using the mean squared error (MSE), which quantifies the deviation from the analytical solution, the latter of which we could derive for the present benchmark case. The obtained MSE of the VQA was then also compared with that of a conservative finite difference method (FDM).

The stationary states obtained for different values of $D_2$ using the VQA are displayed again in Fig. \ref{fig:steady_state_ana_vs_vqa_2}, at the end of the time integration. Our simulations indicate that pronounced hydroxide concentration gradients occur only when the diffusivity ratio $r_D = D_1 / D_2$ exceeds approximately $50$. Under typical operating conditions such extreme contrasts are not expected in hybrid AEM structures. Therefore, we conclude that enhanced membrane degradation resulting from diffusion-driven hydroxide accumulation is unlikely to occur for realistic material parameters. This is the major physical outcome of the present study.

The results obtained by the quantum simulations are a proof of concept that the VQA can solve the diffusion problem numerically. The accuracy of the time evolution in the VQA was bounded below by that of the FDM method, similar to the investigations in ref. \cite{Ingelmann2024}. By comparing the MSE of the VQA and the FDM, we were able to conclude that their performance depends not only on the number of grid points (number of qubits in the VQA) used to resolve the spatial coordinate, but also on the discontinuity of the diffusion constants of the layers of the AEM. More detailed, in the low diffusivity regime, the FDM performance deteriorated, and the VQA performance became almost identical to the FDM. The AEM diffusivity also impacted the computational effort of the optimization part of the VQA, meaning a low diffusivity regime results in faster convergence to the optimal solution, see again Table III. We compared four different classical optimization schemes to find the minimum of the cost function. The surrogate-based optimization scheme by Shaffer et al. \cite{Shaffer2023}, which we adapted to the present problem, did not lead to a better performance for the present application case in comparison to standard algorithms like Nelder-Mead or BFGS schemes. BFGS gave the best results for the ideal statevector case, CMA-ES for the shot-based runs.

Due to the cross-disciplinary nature of the problem considered in this work, the future research directions can be separated into two major categories, namely, (1) physical modeling of the multi-layered AEM and (2) further extensions related to the VQA itself: 
\begin{itemize}
\item To ensure the practical relevance, the two-layer AEM model can be refined by, for instance, considering the AEM diffusivity dependence on the anion concentration field, modeling water concentration within the AEM \cite{Yassin2020} and/or accounting for the electric field distribution \cite{Dekel2018}. This, in turn, will transform the problem from a simple diffusion equation into a system of coupled (nonlinear) one-dimensional partial differential equations, which still can be solved by quantum computing with a moderate number of qubits.

\item Concerning the VQA, the structure of the RPQC has a substantial impact on the high-dimensional parameter landscape of the non-convex optimization problem. The optimization is problem-specific, and thus, complexity estimates for the algorithm as a whole will remain difficult. Therefore, the RPQC has to be chosen according to the diffusion problem, and ideally,  a problem-specific ansatz circuit can lead to drastic improvement towards optimization and necessary fidelity. Apart from that, optimization algorithms that take the structure of PQC into account can be employed, such as natural gradient descent \cite{Stokes2020}, imaginary time evolution \cite{Gacon2021} or curvature-informed gradient-based algorithms \cite{Fitzek2024}. Finally, noisy quantum circuit simulations have to be conducted next, as they can reveal the actual performance of the quantum algorithm on real quantum computing devices. 
\end{itemize}
These efforts are underway and will be reported elsewhere.

\section*{Acknowledgements}
The work of T.G. and P.P. is funded by the European Union (ERC, MesoComp, 101052786). Views and opinions expressed are, however, those of the authors only and do not necessarily reflect those of the European Union or the European Research Council. The work benefited from helpful discussions with Sachin S. Bharadwaj, Sergio Bengoechea, Paul Over, and Katepalli R. Sreenivasan.

\section*{Data Availability}
The data that support the findings of this article are openly available \cite{Gubaev2026}.

\appendix
\section{Derivation of the analytical solution}
\label{app:integrals}

This appendix provides details on the derivation of the analytical solution. To obtain the expansion coefficients $B_{1k}$ of the analytical solution, the integration \eqref{eq:B1k_integral} is carried out. The closed-form expression for this integral is,
\begin{equation}
    \begin{split}
        B_{1k} &=  \frac{1}{G_k}\sum_{j = 1}^m\frac{D_j}{\lambda_k^2}\Bigg[a_j\hat{A}_{jk} + \frac{\lambda_k}{\sqrt{D_j}}(c_{0j} - c_s(\tilde{x}_{j - 1})) \\
        & - \left(a_j\hat{A}_{jk} + \frac{\lambda_k}{\sqrt{D_j}}(c_{0j} - c_s(\tilde{x}_j))\right) \\
        &\times \hat{B}_{jk}\cos\left(\frac{\lambda_k}{\sqrt{D_j}}(\tilde{x}_j - \tilde{x}_{j - 1})\right) \\
        & + \left(-a_j\hat{B}_{jk} + \frac{\lambda_k}{\sqrt{D_j}}(c_{0j} - c_s(\tilde{x}_j))\right) \\
        &\times \hat{A}_{jk}\sin\left(\frac{\lambda_k}{\sqrt{D_j}}(\tilde{x}_j - \tilde{x}_{j - 1})\right)\Bigg]\,,\\
    \end{split}
    \label{eq:app_B1k}
\end{equation}
where $G_k$ are normalization constants. The normalization constants are found by the integration \eqref{eq:Gk_integral}, resulting in the closed-form expression,
\begin{equation}
    \begin{split}
        G_k &= \sum_{j = 1}^m \frac{\sqrt{D_j}}{4\lambda_k}\Bigg[2\hat{A}_{jk}\hat{B}_{jk} \\
        &+ 2\frac{\lambda_k}{\sqrt{D_j}}(\tilde{x}_{j} - \tilde{x}_{j - 1})\left(\hat{A}_{jk}^2 + \hat{B}_{jk}^2\right) \\
        & - 2\hat{A}_{jk}\hat{B}_{jk}\cos\left(2\frac{\lambda_k}{\sqrt{D_j}}(\tilde{x}_{j} - \tilde{x}_{j - 1})\right) \\
        & + \left(\hat{A}_{jk}^2 - \hat{B}_{jk}^2\right)\sin\left(2\frac{\lambda_k}{\sqrt{D_j}}(\tilde{x}_{j} - \tilde{x}_{j - 1})\right)\Bigg]\,.
    \end{split}
    \label{eq:app_Gk}
\end{equation}

\section{Derivation of the VQA minimization problem}
\label{app:vqa}

This appendix lists specifics of the variational formalism. The diffusion operator in \eqref{eq:discrete_weak_formulation_optimization} is discretized at $x = x_j$ by successive application of the central finite difference formula using staggered points $j-\frac{1}{2}$, $j+\frac{1}{2}$ (outer derivative) and collocated points $j$ (inner derivatives) points, namely,
\begin{equation}
    \begin{split}
        & \left[\frac{d}{dx}\left(D(x)\frac{d\tilde c^l}{dx}\right)\right]\bigg |_{x = x_j} \\
        & = \frac{D_{j + \frac{1}{2}}\frac{d\tilde c^l}{dx}\big |_{x = x_{j + \frac{1}{2}}} - D_{j - \frac{1}{2}}\frac{d\tilde c^l}{dx}\big |_{x = x_{j - \frac{1}{2}}}}{\Delta x} \\
        & = \frac{D_{j + \frac{1}{2}}(\tilde c_{j + 1}^l - \tilde c_j^l) - D_{j - \frac{1}{2}}(\tilde c_j^l - \tilde c_{j - 1}^l)}{\Delta x^2} \\
        & = \frac{D_{j - \frac{1}{2}}\tilde c_{j - 1}^l - \left(D_{j - \frac{1}{2}} + D_{j + \frac{1}{2}}\right)\tilde c_j^l + D_{j + \frac{1}{2}}\tilde c_{j + 1}^l}{\Delta x^2}\,.
    \end{split}
    \label{eq:diffusion_operator_discretization}
\end{equation}
Since the transient solution $\tilde c(x, t)$ is subject to homogeneous Dirichlet conditions $\tilde c_0 = \tilde c_{N + 1} = 0$, the first and the last sums in \eqref{eq:vqa_diffusion_operator_discretized} can be combined in $S_\text{DIR}(\tilde c^l)$ as:
\begin{equation}
    \begin{split}
         S_{\text{DIR}}(\tilde c^l) &= \sum_{j = 1}^N \tilde c^l_jD_{j + \frac{1}{2}}\tilde c^l_{j + 1} + \sum_{j = 1}^N \tilde c^l_jD_{j - \frac{1}{2}}\tilde c^l_{j - 1} \\
        & = \left(\underbrace{\tilde c^l_{N}D_{N + \frac{1}{2}}\tilde c^l_{N + 1}}_{= 0} + \sum_{j = 1}^{N - 1} \tilde c^l_jD_{j + \frac{1}{2}}\tilde c^l_{j + 1}\right) \\
        & + \left(\underbrace{\tilde c^l_0D_{\frac{1}{2}}\tilde c^l_1}_{= 0} + \sum_{j = 1}^{N - 1} \tilde c^l_jD_{j + \frac{1}{2}}\tilde c^l_{j + 1}\right) \\
        & = 2\sum_{j = 1}^{N - 1} \tilde c^l_jD_{j + \frac{1}{2}}\tilde c^l_{j + 1}\,.
    \end{split}
    \label{eq:app_S_DIR}
\end{equation}

\end{document}